\documentclass[twocolumn]{aastex62}
\usepackage{lineno}
\pdfoutput=1 
\usepackage{amsmath,amstext}
\usepackage[T1]{fontenc}
\usepackage[figure,figure*]{hypcap}
\graphicspath{{./}{figures/}}

\received{?}
\revised{?}
\accepted{?}
\submitjournal{AAS}

\shorttitle{TOI-1444}
\shortauthors{Dai et al.}

\begin{document}

\title{TKS X: Confirmation of TOI-1444b and a Comparative Analysis of the Ultra-short-period Planets with Hot Neptunes}

\author[0000-0002-8958-0683]{Fei Dai} 
\affiliation{Division of Geological and Planetary Sciences,
1200 E California Blvd, Pasadena, CA, 91125, USA}

\author[0000-0001-8638-0320]{Andrew W. Howard}
\affiliation{Department of Astronomy, California Institute of Technology, Pasadena, CA 91125, USA}

\author[0000-0002-7030-9519]{Natalie M. Batalha}
\affiliation{Department of Astronomy and Astrophysics, University of California, Santa Cruz, CA 95060, USA}

\author[0000-0001-7708-2364]{Corey Beard}
\affiliation{Department of Physics \& Astronomy, The University of California, Irvine, Irvine, CA 92697, USA}

\author[0000-0003-0012-9093]{Aida Behmard}
\affiliation{Division of Geological and Planetary Sciences,
California Institute of Technology,
1200 East California Blvd,Pasadena, CA 91125, USA}

\author[0000-0002-3199-2888]{Sarah Blunt}
\affiliation{Department of Astronomy, California Institute of Technology, Pasadena, CA 91125, USA}

\author{Casey L. Brinkman}
\affiliation{Institute for Astronomy, University of Hawai`i, 2680 Woodlawn Drive, Honolulu, HI 96822, USA}

\author[0000-0003-1125-2564]{Ashley Chontos}
\altaffiliation{NSF Graduate Research Fellow}
\affiliation{Institute for Astronomy, University of Hawai`i, 2680 Woodlawn Drive, Honolulu, HI 96822, USA}

\author{Ian J. M. Crossfield}
\affiliation{Department of Physics \& Astronomy, University of Kansas, 1082 Malott,1251 Wescoe Hall Dr., Lawrence, KS 66045, USA}

\author[0000-0002-4297-5506]{Paul A. Dalba}
\altaffiliation{NSF Astronomy and Astrophysics Postdoctoral Fellow}
\affiliation{Department of Earth and Planetary Sciences, University of California, Riverside, CA 92521, USA}

\author[0000-0001-8189-0233]{Courtney Dressing}
\affiliation{501 Campbell Hall, University of California at Berkeley, Berkeley, CA 94720, USA}

\author[0000-0003-3504-5316]{Benjamin Fulton}
\affiliation{NASA Exoplanet Science Institute/Caltech-IPAC, MC 314-6, 1200 E California Blvd, Pasadena, CA 91125, USA}

\author[0000-0002-8965-3969]{Steven Giacalone}
\affiliation{501 Campbell Hall, University of California at Berkeley, Berkeley, CA 94720, USA}

\author[0000-0002-0139-4756]{Michelle L. Hill}
\affiliation{Department of Earth and Planetary Sciences, University of California, Riverside, CA 92521, USA}

\author[0000-0001-8832-4488]{Daniel Huber}
\affiliation{Institute for Astronomy, University of Hawai`i, 2680 Woodlawn Drive, Honolulu, HI 96822, USA}

\author[0000-0002-0531-1073]{Howard Isaacson}
\affiliation{{Department of Astronomy,  University of California Berkeley, Berkeley CA 94720, USA}}
\affiliation{Centre for Astrophysics, University of Southern Queensland, Toowoomba, QLD, Australia}

\author[0000-0002-7084-0529]{Stephen R. Kane}
\affiliation{Department of Earth and Planetary Sciences, University of California, Riverside, CA 92521, USA}

\author[0000-0001-8342-7736]{Jack Lubin}
\affiliation{Department of Physics \& Astronomy, The University of California, Irvine, Irvine, CA 92697, USA}

\author[0000-0002-7216-2135]{Andrew Mayo}
\affiliation{501 Campbell Hall, University of California at Berkeley, Berkeley, CA 94720, USA}

\author[0000-0003-4603-556X]{Teo Mo\v{c}nik}
\affiliation{Gemini Observatory/NSF's NOIRLab, 670 N. A'ohoku Place, Hilo, HI 96720, USA}

\author[0000-0001-8898-8284]{Joseph M. Akana Murphy}
\altaffiliation{NSF Graduate Research Fellow}
\affiliation{Department of Astronomy and Astrophysics, University of California, Santa Cruz, CA 95060, USA}

\author[0000-0003-0967-2893]{Erik A. Petigura}
\affiliation{Department of Physics \& Astronomy, University of California Los Angeles, Los Angeles, CA 90095, USA}

\author[0000-0002-7670-670X]{Malena Rice}
\affiliation{Department of Astronomy, Yale University, New Haven, CT 06511, USA}

\author[0000-0003-0149-9678]{Paul Robertson}
\affiliation{Department of Physics \& Astronomy, University of California Irvine, Irvine, CA 92697, USA}

\author[0000-0001-8391-5182]{Lee Rosenthal}
\affiliation{Department of Astronomy, California Institute of Technology, Pasadena, CA 91125, USA}

\author[0000-0001-8127-5775]{Arpita Roy}
\affiliation{Space Telescope Science Institute, 3700 San Martin Drive, Baltimore, MD 21218, USA}
\affiliation{Department of Physics and Astronomy, Johns Hopkins University, 3400 N Charles St, Baltimore, MD 21218, USA}

\author[0000-0003-3856-3143]{Ryan A. Rubenzahl}
\altaffiliation{NSF Graduate Research Fellow}
\affiliation{Department of Astronomy, California Institute of Technology, Pasadena, CA 91125, USA}

\author[0000-0002-3725-3058]{Lauren M. Weiss}
\affiliation{Institute for Astronomy, University of Hawai`i, 2680 Woodlawn Drive, Honolulu, HI 96822, USA}

\author[0000-0002-4290-6826]{Judah Van Zandt}
\affil{Department of Physics \& Astronomy, University of California Los Angeles, Los Angeles, CA 90095, USA}

\author[0000-0002-5627-5471]{Charles Beichman}
\affiliation{NASA Exoplanet Science Institute/Caltech-IPAC, MC 314-6, 1200 E California Blvd, Pasadena, CA 91125, USA}

\author[0000-0002-5741-3047]{David Ciardi}
\affiliation{NASA Exoplanet Science Institute/Caltech-IPAC, MC 314-6, 1200 E California Blvd, Pasadena, CA 91125, USA}

\author[0000-0001-6588-9574]{Karen A.\ Collins}
\affiliation{Center for Astrophysics \textbar \ Harvard \& Smithsonian, 60 Garden Street, Cambridge, MA 02138, USA}

\author{Erica Gonzales}
\affiliation{Department of Astronomy and Astrophysics, University of California, Santa Cruz, CA 95060, USA}

\author[0000-0002-2532-2853]{Steve~B.~Howell}
\affiliation{NASA Ames Research Center, Moffett Field, CA 94035, USA}

\author[0000-0001-7233-7508]{Rachel~A.~Matson}
\affiliation{U.S. Naval Observatory, Washington, D.C. 20392, USA}

\author[0000-0003-0593-1560]{Elisabeth C. Matthews}
\affiliation{Observatoire de l’Universit\'e de Gen\`eve, Chemin Pegasi 51, 1290 Versoix, Switzerland}

\author[0000-0001-5347-7062]{Joshua E. Schlieder}
\affiliation{NASA Goddard Space Flight Center, 8800 Greenbelt Rd., Greenbelt, MD 20771, USA}

\author[0000-0001-8227-1020]{Richard P. Schwarz}
\affiliation{Patashnick Voorheesville Observatory, Voorheesville, NY 12186, USA}

\author[0000-0003-2058-6662]{George R. Ricker}
\affiliation{Department of Physics and Kavli Institute for Astrophysics and Space Research, Massachusetts Institute of Technology, Cambridge, MA 02139, USA}

\author[0000-0001-6763-6562]{Roland Vanderspek}
\affiliation{Department of Physics and Kavli Institute for Astrophysics and Space Research, Massachusetts Institute of Technology, Cambridge, MA 02139, USA}

\author[0000-0001-9911-7388]{David W. Latham}
\affiliation{Center for Astrophysics | Harvard \& Smithsonian, 60 Garden St, Cambridge, MA 02138, USA}

\author[0000-0002-6892-6948]{Sara Seager}
\affiliation{Department of Physics and Kavli Institute for Astrophysics and Space Research, Massachusetts Institute of Technology, Cambridge, MA
02139, USA}
\affiliation{Department of Earth, Atmospheric and Planetary Sciences, Massachusetts Institute of Technology, Cambridge, MA 02139, USA}
\affiliation{Department of Aeronautics and Astronautics, MIT, 77 Massachusetts Avenue, Cambridge, MA 02139, USA}

\author[0000-0002-4265-047X]{Joshua N. Winn}
\affiliation{Department of Astrophysical Sciences, Princeton University, 4 Ivy Lane, Princeton, NJ 08544, USA}

\author[0000-0002-4715-9460]{Jon M. Jenkins}
\affiliation{NASA Ames Research Center, Moffett Field, CA 94035, USA}

\author[0000-0002-4715-9460]{Douglas A. Caldwell}
\affiliation{NASA Ames Research Center, Moffett Field, CA 94035, USA}

\author{Knicole D. Colon}
\affiliation{NASA Goddard Space Flight Center, Exoplanets and Stellar Astrophysics Laboratory (Code 667), Greenbelt, MD 20771, USA}

\author[0000-0003-2313-467X]{Diana Dragomir}
\affiliation{Department of Physics and Astronomy, University of New Mexico, 1919 Lomas Blvd NE, Albuquerque, NM 87131, USA}

\author[0000-0003-2527-1598]{Michael B. Lund}
\affiliation{NASA Exoplanet Science Institute/Caltech-IPAC, MC 314-6, 1200 E California Blvd, Pasadena, CA 91125, USA}

\author{Brian McLean}
\affiliation{Space Telescope Science Institute, 3700 San Martin Drive, Baltimore, MD, 21218, USA}

\author{Alexander Rudat}
\affiliation{Department of Physics and Kavli Institute for Astrophysics and Space Research, Massachusetts Institute of Technology, Cambridge, MA 02139, USA}

\author[0000-0002-1836-3120]{Avi Shporer}
\affiliation{Department of Physics and Kavli Institute for Astrophysics and Space Research, Massachusetts Institute of Technology, Cambridge, MA 02139, USA}



\begin{abstract}
\noindent We report the discovery of TOI-1444b, a 1.4-$R_\oplus$ super-Earth on a  0.47-day orbit around a Sun-like star discovered by {\it TESS}. Precise radial velocities from Keck/HIRES confirmed the planet and constrained the mass to be $3.87 \pm 0.71 M_\oplus$. The RV dataset also indicates a possible  non-transiting,  16-day planet ($11.8\pm2.9M_\oplus$). We report a tentative detection of phase curve variation and secondary eclipse of TOI-1444b in the {\it TESS} bandpass. TOI-1444b joins the growing sample of 17 ultra-short-period planets with well-measured masses and sizes, most of which are compatible with an Earth-like composition. We take this opportunity to examine the expanding sample of ultra-short-period planets ($<2R_\oplus$) and contrast them with the newly discovered sub-day ultra-hot Neptunes ($>3R_\oplus$, $>2000F_\oplus$  TOI-849 b, LTT9779 b and K2-100). We find that 1) USPs have predominately Earth-like compositions with inferred iron core mass fractions of 0.32$\pm$0.04; and have masses below the threshold of runaway accretion ($\sim 10M_\oplus$), while ultra-hot Neptunes are above the threshold and have H/He or other volatile envelope. 2) USPs are almost always found in multi-planet system consistent with a secular interaction formation scenario; ultra-hot Neptunes ($P_{\rm orb} \lesssim$1 day) tend to be ``lonely' similar to longer-period hot Neptunes($P_{\rm orb}$1-10 days) and hot Jupiters. 3) USPs occur around solar-metallicity stars while hot Neptunes prefer higher metallicity hosts. 4) In all these respects, the ultra-hot Neptunes show more resemblance to hot Jupiters than the smaller USP planets, although ultra-hot Neptunes are rarer than both USP and hot Jupiters by 1-2 orders of magnitude.
\end{abstract}


\section{Introduction}
The widely-used term ``ultra-short-period'' planets (USP) usually refers to terrestrial planets that orbit their host stars in less than 1 day. The current record holders have an orbital period of just 4 hours --- on the verge of tidal disruption \citep[e.g. KOI-1843.03, K2-137b;][]{Rappaport,Smith}. USP are found around $\approx$0.5\% of Sun-like stars while their radii are generally smaller than 2$R_\oplus$ \citep{USP}. As a group, USPs have been the Rosetta Stone for probing the composition of terrestrial planets. A true Earth analog has a radial velocity (RV) semi-amplitiude of just 9~cm s$^{-1}$ which is beyond the reach of current state-of-art spectrographs. With a much greater gravitational pull on the host star, a USP usually has a semi-ampltiude that is of order several m s$^{-1}$, hence above the limit of both instrumental uncertainty and typical stellar activity jitter. The short orbital periods of USP also provide a strong timescale contrast when compared to the host star rotation period, making it easier to disentangle the planetary signal from stellar activity in RV analysis. Moreover, USPs are so strongly irradiated that any primordial H/He atmosphere has probably been eroded by photoevaporation \citep{USP,Lundkvist,Lopez2017}. One can directly constrain the composition of the rocky cores without worrying about the degeneracy caused by a thick atmosphere. Finally, USP planets are amenable to phase curve variation and secondary eclipse studies. The resultant albedo, phase offset and day-night temperature contrast directly probes the planet's surface composition \citep{Demory,LHS3844}.

The extreme orbits of USP planets dare theorists to come up with an explanation. Many USP orbits are so close to their hosts that they are within the dust sublimation radius \citep[$a/R_\star \sim 8$ for Sun-like stars,][]{Isella} or even within what would have been the radius of the once younger host stars. It appears extremely unlikely that USPs formed on their current-day orbits. An early proposal is that USPs may be the tidally disrupted cores of hot Jupiters that likely formed further out in the disk before migrating to their current day orbit \citep[e.g.][]{Jackson}. This idea is now disfavored because hot Jupiters are observed to preferentially occur around metal-rich host stars \citep{Fischer2005}; whereas \citet{Winn2017} showed USP hosts have a statistically distinct, more solar-like metallicity distribution (Fig. \ref{fig:metallicity}). The USP hosts and hot Jupiter hosts also display distinct distributions in measured kinematics and inferred stellar age \citep{Hamer2019,Hamer2020}: hot Jupiter hosts tend to be younger than USP hosts and other field stars. Another theory for USP formation involving secular interaction now seems more promising \citep{Petrovich,Pu}. In short, USPs initially formed on orbits of a few days similar to many other {\it Kepler} sub-Neptunes before secular interaction with other planets launched them into eccentric, inclined orbits that eventually tidally shrunk to the current-day orbits. The observed high mutual inclinations and large orbital period ratio with neighboring planets lend support to this theory \citep{Dai2018,Steffen2016}. However, other theories involving host star oblateness \citep{Li_USP}, obliquity tides \citep{Millholland_USP} or a more distant companion \citep{Becker_usp} challenge the secular theory as the unique narrative for USP formation.

Given the observational opportunities and theoretical challenges USPs offer, there has been a growing interest \citep[e.g.][]{superpigs,Shporer,Cloutier,Espinoza} in studying the USPs especially using the fresh sample of bright USP hosts discovered by the {\it TESS} mission \citep{Ricker}. \citet{Dai2019} performed a homogeneous analysis of 10 well characterized USP planets. The results suggest that USP planets are predominantly rocky bodies that are consistent with an Earth-like composition. The sample size has now increased to 17. Meanwhile, a few cases of ``Hot Neptune Desert Stragglers'' i.e. planets with Neptune radius or larger but on strongly irradiated, sometimes sub-day orbits \citep[K2-100b, TOI-849b and LTT 9779b][]{Mann100,Armstrong849,Jenkins} have been reported. Such close-in orbits were previously considered forbidden for Neptune-size planet in the sense that strong photoevaporation should quickly strip them down to rocky cores. Yet three such planets have been confirmed (Table 3 and Fig. \ref{fig:insolation}) and their mass and radius suggest substantial H/He atmosphere or icy envelopes despite the strong insolation they receive. In this work, we report the discovery and detailed characterization of TOI-1444b; and we also make use of the opportunity to contrast the USPs and ultra-hot Neptunes in host star properties, orbital architecture and formation pathways. 

The paper is structured as follows. Section 2 presents the characterization of the TOI-1444 host star. Section 3 outlines the adaptive optics imaging of the star that rules out close stellar companions. In Section 4, we present photometric analysis of the system from transit modeling, phase curve analysis to the search of additional planets. In Section 5, we present the RV follow-up of TOI-1444 and the resultant constraint on the planet mass. Section 6 discusses the updated list of USP planets and their relation to the hot Neptunes. Finally, we summarize the results of this work in Section 7.

\section{Host Star Properties} \label{sec:host}

To derive the spectroscopic parameters ($T_{\rm eff}$, log $g$ and [Fe/H]) of TOI-1444, we obtained a high-SNR iodine-free spectra with High Resolution
Echelle Spectrometer on the 10m Keck I telescope (Keck/HIRES) on UT Aug 17 2020. We applied the {\tt SpecMatch-Syn} pipeline\footnote{\url{https://github.com/petigura/specmatch-syn}} \citep{CKS1} to this spectrum. {\tt SpecMatch-Syn} makes use of interpolation based on a precomputed grid of theoretical stellar spectra at discrete $T_{\rm eff}$, [Fe/H], log$~g$ values \citep{Coelho}. Broadening effects due to stellar rotation and macroturbulence are convolved with the model spectrum using the prescription of \citet{Hirano}. From our experience with HIRES, instrumental broadening is well described by a Gaussian function with a FWHM of 3.8 pixels. Such an instrumental profile usually captures the shapes of observed telluric lines. {\tt SpecMatch-Syn} fits the best spectroscopic parameters to five $\sim400$ Å spectral segments separately before taking the weighted average. We also correct for known systematic biases of {\tt SpecMatch-Syn} e.g. its higher surface gravity log~$g$  ($\sim$ 0.1 dex) for earlier-type stars when calibrated with the asteroseismic sample of \citet{Huber2013}. For more detail on {\tt SpecMatch-Syn}, please refer to \citet{Petigura_thesis}.

We then combined the Gaia parallax information
\citep{Gaia} and our spectroscopic parameters to derive the stellar parameters. We used the parallax of $7.9779\pm0.0088$ mas reported by Gaia EDR3 \citep{EDR3}. This offers an independent constraint on the stellar radius using Stefan-Boltzmann Law. Given the effective temperature from spectroscopic and the $K$-band magnitude which suffers less from extinction and the parallax information from Gaia, one can calculate the radius of the star. This is implemented in the  {\tt Isoclassify} package \citep{Huber} which combines the spectroscopic parameters and the parallaxes into an isochronal fitting. The measured stellar properties were fitted with the models of MESA Isochrones \& Stellar Tracks \citep[MIST,][]{MIST}. Table 1 summarizes the posterior distribution of various stellar parameters. We caution the readers that  {\tt Isoclassify} does not account for systematic errors between different theoretical model grids which can lead systematic uncertainties of $\sim2\%$ on $T_{\rm eff}$, $\sim4\%$ on $M_{\star}$ and $\sim5\%$ on $R_{\star}$ \citep{Tayar}.

\begin{deluxetable*}{lcc}
\tablecaption{Stellar Parameters of TOI-1444} \label{tab:stellar_para}
\tablehead{
\colhead{Parameters} & \colhead{Value and 68.3\% Confidence Interval} & \colhead{Reference}}

\startdata
TIC ID  & 258514800 & A\\
R.A.  & 20:21:53.98 & A\\
Dec.  & 70:56:37.34 & A\\
V (mag) & 10.936 $\pm$ 0.009& A\\
K (mag) & 9.061 $\pm$ 0.021& A\\
Effective Temperature $T_{\text{eff}} ~(K)$ & $5430\pm90$ & B \\
Surface Gravity $\log~g~(\text{dex})$ &$4.49 \pm 0.03$& B \\
Iron Abundance $[\text{Fe/H}]~(\text{dex})$ &$0.13 \pm 0.06$& B \\
Rotational Broadening $v~\text{sin}~i$ ~(km~s$^{-1}$) &$<2     $& B \\
Stellar Mass $M_{\star} ~(~M_{\odot})$ &$0.934\pm0.038$& B \\
Stellar Radius $R_{\star} ~(R_{\odot})$ &$0.907\pm0.016$& B \\
Stellar Density $\rho_\star$ (g cm$^{-3}$) &$1.24\pm0.10$&B \\
Limb Darkening u$_1$ & $0.46\pm0.28$& B\\
Limb Darkening u$_2$ & $0.21\pm0.17$& B\\
Distance $d$ (pc)&$125\pm2$&B\\
\enddata
\tablecomments{A:TICv8; B: this work.}
\end{deluxetable*}

\section{Adaptive Optics Imaging}
\begin{figure}
\includegraphics[width = 1.\columnwidth]{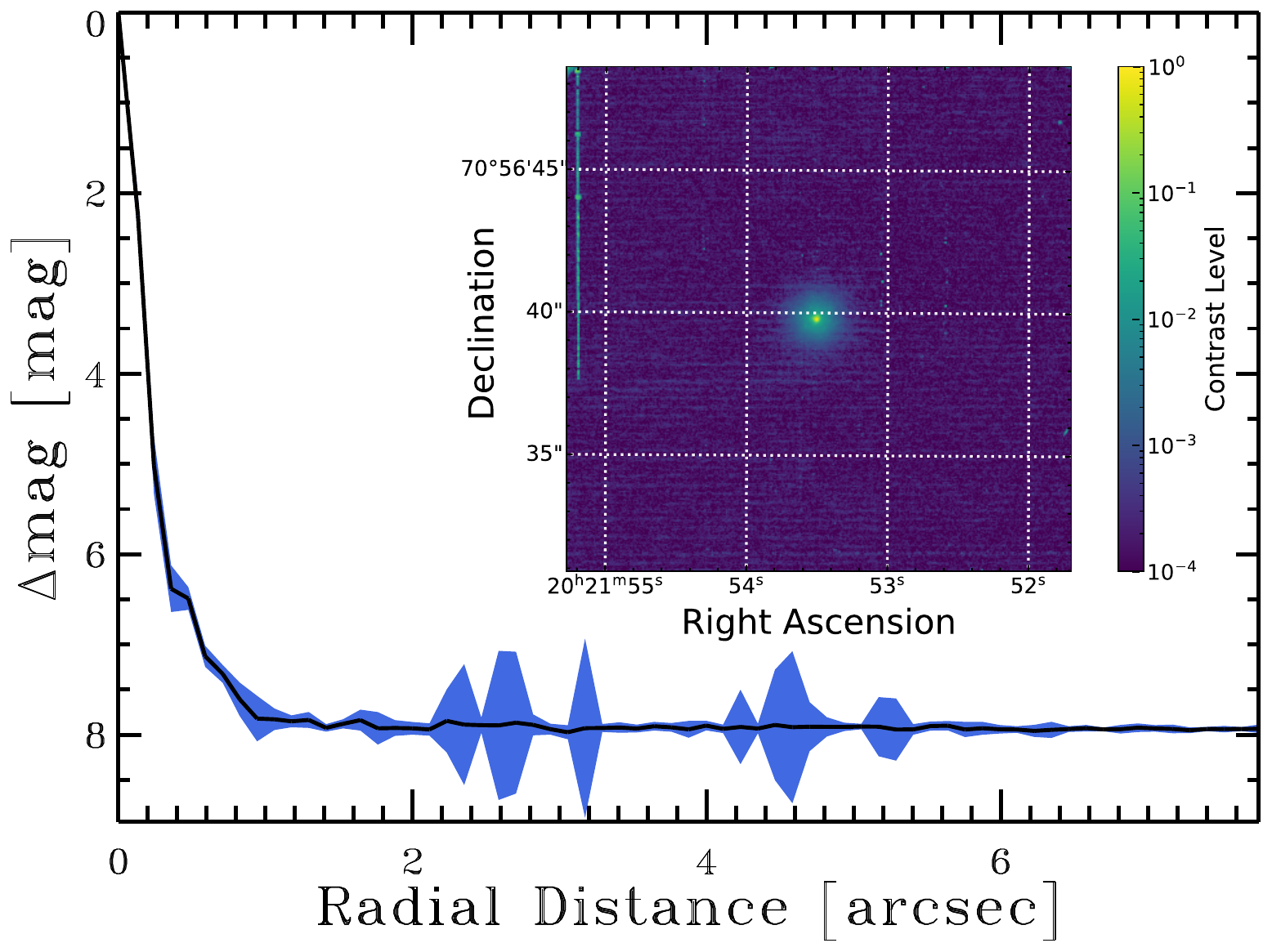}
\caption{The contrast curve as function of radial separation from TOI-1444 using the AO Imaging from Gemini-NIRI. The inset is the image of target star.}
\label{fig:AO}
\end{figure}

We searched for close visual companions of TOI-1444 with Adaptive Optics (AO) imaging. Close visual companions can bias the measured radius of a planet, or even be the source of false positives if the companion is itself an eclipsing binary. Data were collected on UT Dec 04 2019 with Gemini/NIRI \citep{Hodapp}. We collected 9 frames each with exposure time 8.2s in the Br$\gamma$ filter, and dithered the frame between each image by $\sim$1.3'' in a grid pattern. The dithered science frames themselves are median combined to create a sky background frame. We reduced the data using a custom IDL code, which removes bad pixels, corrects for the sky background, flat-fields the data, and aligns and co-adds the individual images. We searched for companions by eye, and did not identify any additional sources within the field of view, which extends to at least 8'' from the star in all directions. To assess the sensitivity of these images to companions, we injected several faint, fake companions into the data and scaled their brightness such that they would be detected at 5$\sigma$. The 5$\sigma$ sensitivity is shown in Fig. \ref{fig:AO} as a function of radius, along with a thumbnail image of the target.

\section{Photometry} \label{sec:photo}

The {\it TESS} \citep{Ricker} mission observed TOI-1444 in Sectors 15, 16, 17, 18, 19, 22, 24, 25 and 26 from UT Aug 15 2019 to July 04 2020. We downloaded the PDC\_SAP light curves \citep{Stumpe2012,Stumpe2014} of all sectors from the Mikulski Archive for Space
Telescopes website\footnote{\url{https://archive.stsci.edu}}. We removed all data points with a non-zero Quality Flag i.e. those suffering from cosmic rays or other known systematic issues. 

\subsection{Stellar Rotation Period}

The {\it TESS} team reported one object of interest TOI-1444.01 with an orbital period of 0.47 day.  Beginning with the transit parameters reported on the ExoFOP website\footnote{\url{https://exofop.ipac.caltech.edu}.}, we removed the data points taken within the transit window of TOI-1444.01. Using the resultant, out-of-transit light curve, we tried to measure the stellar rotation period of TOI-1444 using both Lomb-Scargle periodogram \citep{Lomb1976,Scargle1982} and auto-correlation function \citep[ACF, ][]{McQuillan2014}. These two methods basically look for any quasi-periodic flux modulations that may be attributed to stellar rotation coupled with surface magnetic activity. Neither the periodogram nor the ACF detected a signal above 1\% False Positive rate. Moreover, the highest peaks reported by these two methods did not agree. Therefore we could not measure the rotation period of the host from the flux variation seen in {\it TESS} light curve. The weak flux variation (standard deviation of about 1200 ppm for 2-min cadence over 300-day baseline) may be the result of a slow rotation and/or low stellar activity. Rotational broadening of TOI-1444 was not detectable in our spectrum above other broadening effects. Following \citet{Petigura_thesis} we place an upper limit of  $v~\text{sin}~i~<2$km~s$^{-1}$. The proxy for chromospheric activity in the Ca II H\&K lines $S_{HK}$ is $0.145 \pm 0.004$. This is slightly lower than the median $S_{HK}$ of stars of similar B-V color \citep[ 0.146,][]{Isaacson}.

\subsection{Search for Additional Transiting Planets}

 TOI-1444b was initially detected by the TESS Science Processing Operations Center \citep[SPOC,][]{jenkinsSPOC2016} in a transit search of Sectors 15 and 16 that occurred. The 0.47-day signal was detected at a 7.6 $\sigma$ level with an adaptive, noise-compensating matched filter \citep{2010SPIE.7740E..0DJ}; passed all the diagnostic tests performed and published in the resulting Data Validation reports. The vetting included tests for eclipsing binaries, such as an odd and even depth variation test,  a secondary eclipse test, and a ghost diagnostic test. The difference imaging centroid test showed that the source of the transit signature was consistent with the target star TIC 258514800, but could not exclude nearby stars in the TIC catalog. The signal strength grew as additional observations were collected by {\it TESS}, and the difference imaging centroid test located the source within 3.6'' once the full set of observations were completed in Sector 26. 
 
We searched the {\it TESS} light curve for any additional transit signals particularly that around the 16-day periodicity of the candidate planetary signal seen in the radial velocity dataset (Section \ref{sec:rv}). We first removed the data points within the transit window of TOI-1444.01. We then fitted a cubic spline of length 1.5 days to detrend any long-term stellar or instrumental flux variation. We applied the Box-Least-Squares algorithm \citep[BLS,][]{Kovac2002} to the resultant light curve. Our BLS pipeline is implemented in {\tt C++} and has yielded a number of planet discoveries including other ultra-short-period planets e.g. K2-131b \citep{DaiK2} and K2-141b \citep{Barragan141}. We followed the suggested improvement of BLS as outlined in \citet{Ofir2014}. This involves using a nonlinear frequency grid given the theoretical scaling of transit duration with orbital period for stars of a certain mean density. We also adopted the signal detection efficiency (SDE)  defined in \citet{Ofir2014} to quantify the significance of a BLS signal. In short, SDE is the local variation of the BLS spectrum normalized by the local standard deviation. This helps to remove any period-related systematics.

We recovered TOI-1444.01 with a SDE =21. However, we did not detect any additional transiting signal with a SDE larger than 5. Visual inspection of the phase-folded light curve shows that none of the top candidate signals has a transit-like shape. We also did not detect any convincing single-transit events visually. Notably, there is no transiting signal consistent with the 16-day orbital period of the candidate planetary signal in the RV dataset. No additional transiting signal was found by the SPOC pipeline either.

\subsection{Transit Modeling}\label{sec:transit}
We modeled the transit light curve of TOI-1444.01 or TOI-1444b to constrain its transit parameters. We started from the transit ephemeris reported by the {\it TESS} team. We used the {\tt Python} package {\tt Batman} \citep{Kreidberg2015} to generate the model transit light curves. We also imposed a prior on the host star mean density using the result in Table 1. This precise prior on mean stellar density helps to break some of the degeneracy in modeling transit morphology and often lead to improved transit parameters \citep{Seager}. We adopted a quadratic limb darkening profile. We imposed Gaussian priors (width of 0.3) on the limb-darkening coefficients $u_1$ and $u_2$. We queried the EXOFAST\footnote{\url{astroutils.astronomy.ohio-state.edu/exofast/limbdark.shtml}.}  \citep{Eastman2013} for the theoretical limb darkening given the spectroscopic parameters of TOI-1444 ($u_1$= 0.48, $u_2$ = 0.22). We also adopted the efficient sampling reparameterization of limb darkening coefficients proposed by \citet{Kipping}. The other parameters in our transit model are the orbital period $P_{\text{orb}}$, the time of conjunction $T_{\text{c}}$, the planet-to-star radius ratio $R_{\text{p}}/R_\star$, the scaled orbital distance $a/R_\star$ and the impact parameter $b\equiv a\cos i/R_\star$. 

We first fitted all transits of TOI-1444b globally. We found the best-fit solution using the Levenberg-Marquardt method specifically using that implemented in {\tt Python} package {\tt lmfit}. Using this global fit as a template, we then fitted each individual transit allowing the mid-transit time to vary freely. The resultant transit times do not show quasi-sinusoidal variation as one would expect in the case of transit-timing variations. We did not detect any prominent periodicity; instead it is consistent with a linear ephemeris which we assume in subsequent analysis.

\begin{figure}
\includegraphics[width = 1.\columnwidth]{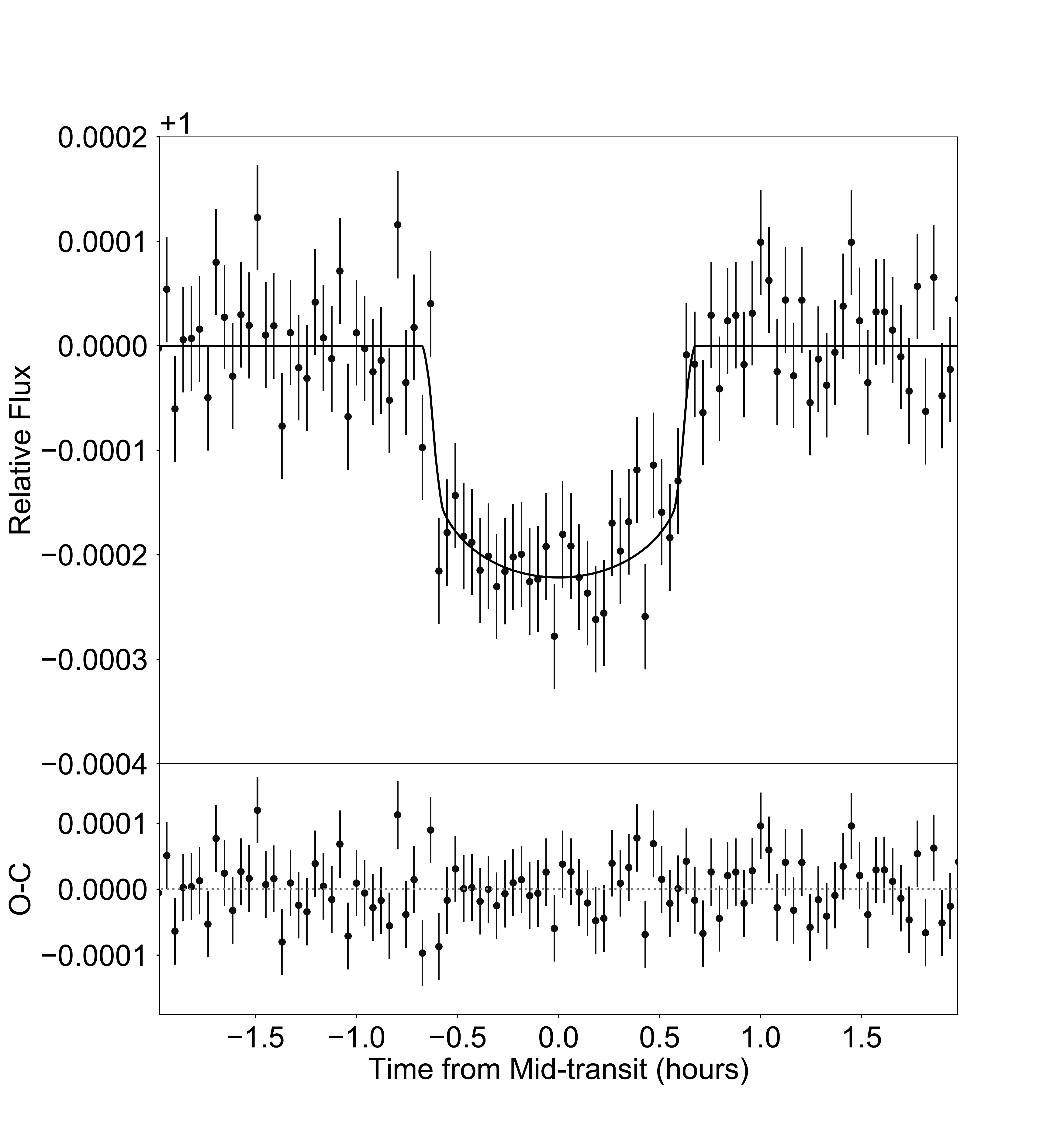}
\caption{Phase folded and binned 9 sectors of {\it TESS} light curve and the best fit transit model for planet b.}
\label{fig:transit}
\end{figure}

We sampled the posterior distribution of transit parameters with the affine-invariant Markov Chain Monte Carlo method as implemented in {\tt Python} package {\tt emcee} \citep{emcee}. We used 128 walkers starting from the maximum likelihood solution found by  {\tt lmfit}. With 50000 links, the Gelman-Rubin potential scale reduction factor reduced to below 1.01 suggesting convergence of the sampling process. We summarize the resultant posterior distribution in Table 2. Fig. \ref{fig:transit} shows the phase-folded and binned {\it TESS} light curve and the best-fit transit model.

\subsection{Phase Curve and Secondary Eclipse}
\begin{figure}
\begin{center}
\includegraphics[width = 1.\columnwidth]{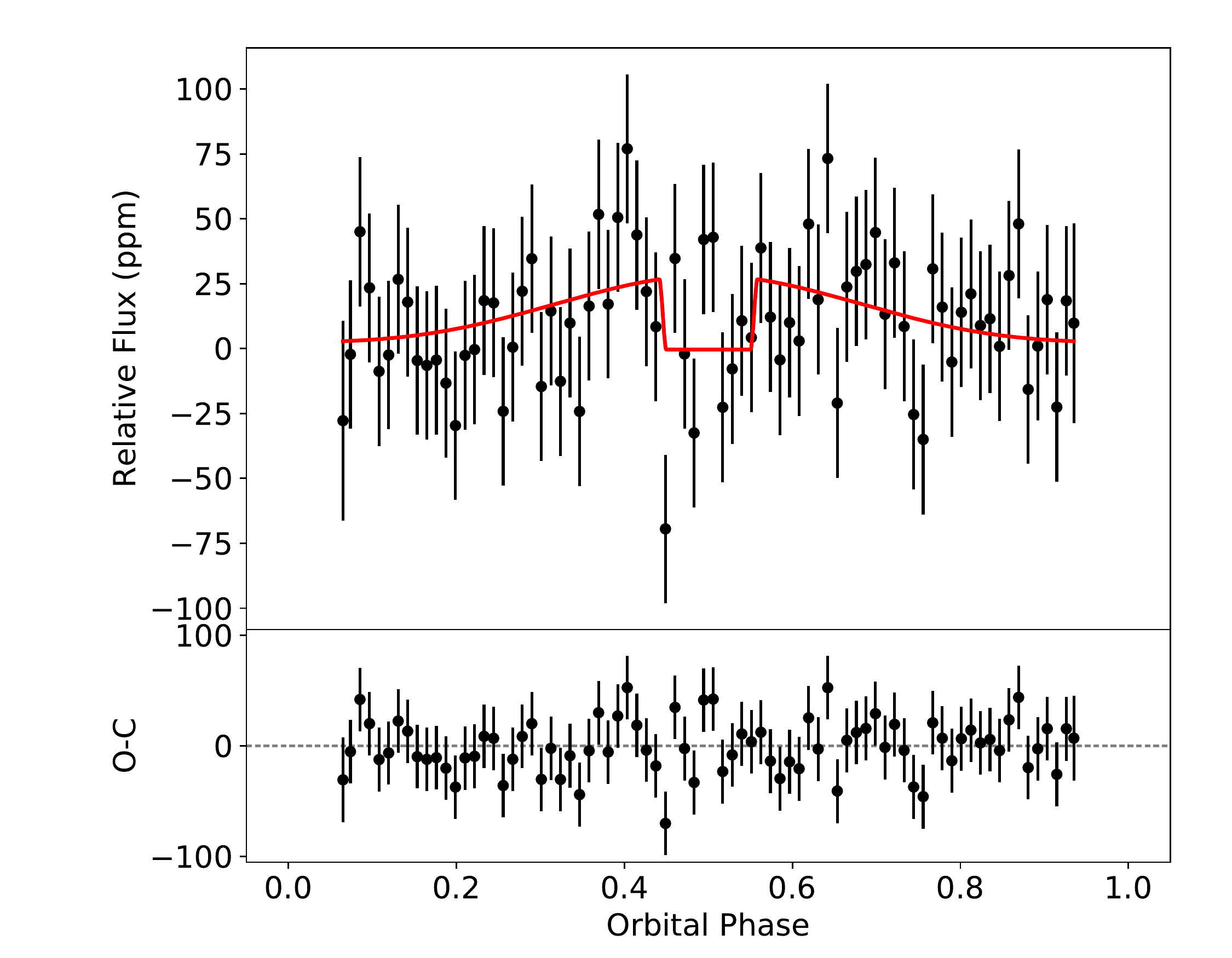}
\caption{The phase-folded and binned out-of-transit flux variation of TOI-1444 b in {\it TESS} observation,  The red curve shows the best fit phase curve and secondary eclipse model.}
\label{fig:phase}
\end{center}
\end{figure}

We looked for any phase curve variation and secondary eclipse of TOI-1444b in the {\it TESS} band. Again we removed data taken within the transit window of TOI-1444b. We then detrended any long-term stellar variation or systematic effect using the procedure outlined in \citet{Sanchis78}. In short, a linear function in time whose width is at least one orbital period is fitted to remove local flux variation longer than the orbital period. In practice, we tried 1$\times$, 2$\times$ or 3$\times P_{\rm orb}$, and found nearly identical results. We also compared PDC\_SAP and SAP versions of {\it TESS} photometry; and found no substantial difference. Fig. \ref{fig:phase} shows the phase curve variation from all {\it TESS} sectors of PDC\_SAP photometry.

We then modeled the phase curve variation and secondary eclipse simultaneously. We modeled the secondary eclipse using {\tt Batman}. We fixed the transit parameters using the best-fit primary transit solution and changed the limb darkening coefficients to 0. The phase curve variation was modeled as a Lambertian disk \citep[e.g.][]{Demory}. The parameters in this model include the depth of secondary eclipse $\delta_{\rm sec}$, the amplitude of illumination effect $A_{\rm ill}$, the time of secondary eclipse $t_{\rm sec}$ and the phase offset of the illumination effect $\theta_{\rm ill}$. To constrain the posterior distribution, we similarly ran an MCMC analysis with {\tt emcee} \citep{emcee}. We found that both $\delta_{\rm sec} = 27\pm12$ ppm and $A_{\rm ill} = 24 \pm 10$ ppm showed marginal detection of no more than 2.5 $\sigma$. The fact that $\delta_{\rm sec}$ are similar to $A_{\rm ill}$ in amplitude even though they are fitted separately boosts our confidence in these detections. $t_{\rm sec}$ centers around the half-way from mid-transit although with substantial errorbar $t_{\rm sec} = 0.236 \pm 0.019$ day or 0.502$\pm0.040$ in orbital phase. Again this is expected given the very short tidal circularization timescale of a planet on such a short-period orbit. The phase curve offset $\theta_{\rm ill} = 16\pm27^{\circ}$ shows a very weak preference for an eastward offset.

\begin{figure}
\begin{center}
\includegraphics[width = 1.\columnwidth]{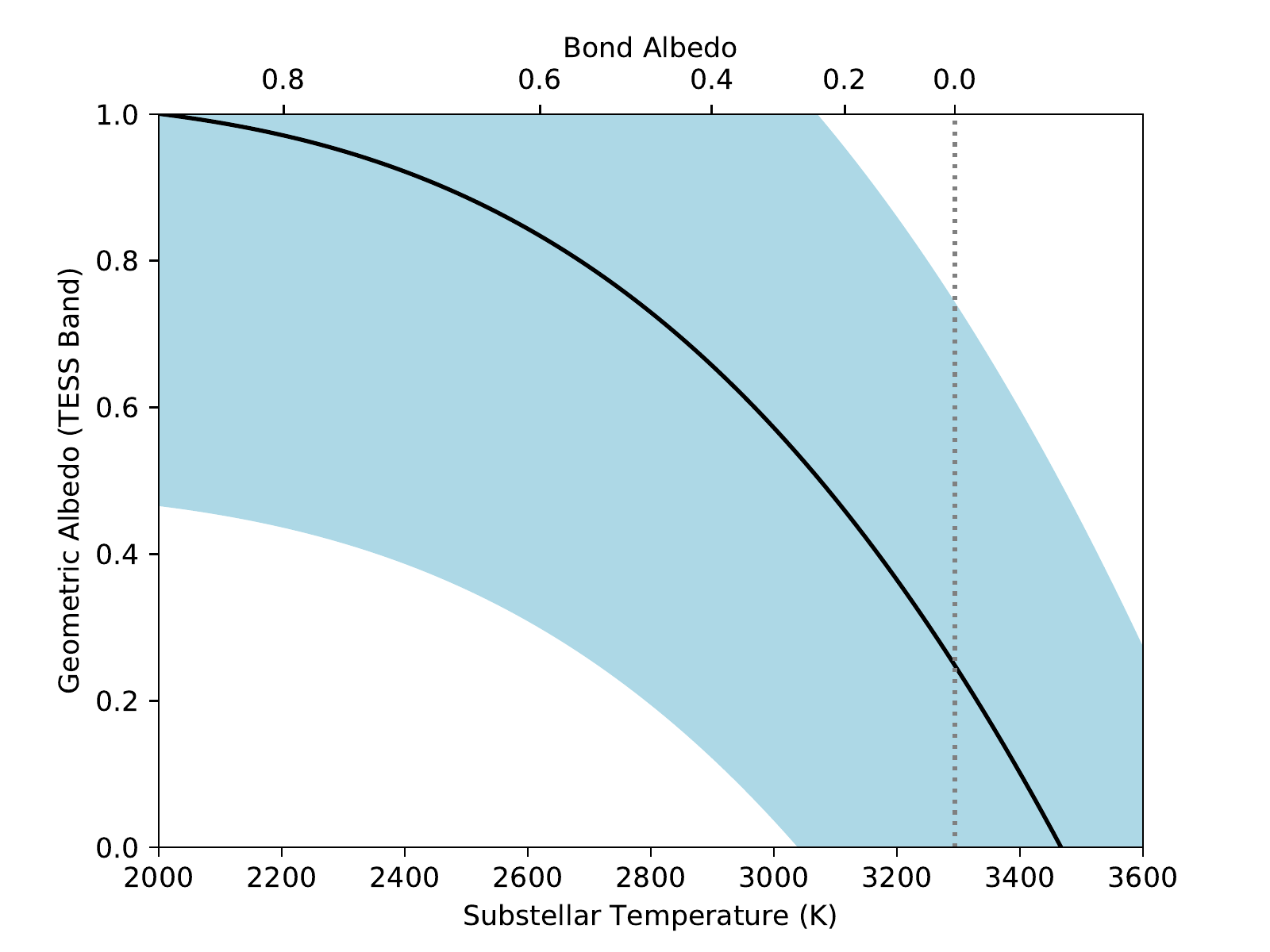}
\caption{The geometric albedo (reflected light) versus the Bond albedo (thermal emission) degeneracy that can reproduce the phase curve and secondary eclipse observation in {\it TESS} band. Blue-shaped region shows the 1-$\sigma$ confidence interval. }
\label{fig:albedo}
\end{center}
\end{figure}

The {\it TESS} passband (600-1000 nm) extends substantially into the infrared that one may expect to see thermal emission from the planet as well as reflected star light. However, with only one band, we could not break the degeneracy between reflected stellar light and thermal emission from the planet. Fig. \ref{fig:albedo} captures this degeneracy in the Bond Albedo versus {\it TESS} band Geometric Albedo plane. The blue-shaded region shows the 68\% confidence interval. Given how wide the confidence interval is, we refrain from making a strong interpretation of the marginal 2.5 $\sigma$ phase curve detection in  {\it TESS} band. Phase curve observation in the infrared would consolidate the phase curve variation and may reveal the surface properties of TOI-1444b.

\subsection{Ground-based Follow-up}
 We observed TOI-1444 on May 02 2020 during the transit window of planet b as predicted by the SPOC pipeline analysis of TESS Sectors 15 and 16. The observation was carried out in the Pan-STARSS $z$-short band from the Las Cumbres Observatory Global Telescope \citep[LCOGT;][]{Brown:2013} 1.0\,m network node at the McDonald Observatory. The 4096x4096 LCO SINISTRO cameras have an image scale of $0\farcs389$ per pixel, providing a field of view of $26\arcmin\times26\arcmin$. The standard LCOGT BANZAI pipeline \citep{McCully2018} was used to calibrate the images, and the photometric data were extracted with the AstroImage (AIJ) software package \cite[][]{Collins:2017}.

Since the $\sim 200$ ppm event detected by the SPOC pipeline is generally too shallow to detect with ground-based observations, we checked for a faint nearby eclipsing binary (NEB) that could be contaminating the SPOC photometric aperture. To account for possible contamination from the wings of neighboring star PSFs, we searched for NEBs at the positions of Gaia DR2 stars out to $2\farcm5$ from the target star. If fully blended in the SPOC aperture, a neighboring star that is fainter than the target star by 9.3 magnitudes in TESS-band could produce the SPOC-reported flux deficit at mid-transit (assuming a 100\% eclipse). To account for possible delta-magnitude differences between TESS-band and Pan-STARSS $z$-short band, we included an extra 0.5 magnitudes fainter (down to \textit{TESS}-band magnitude 20.0). We visually compared the light curves of the 85 nearby stars that meet our search criteria with models that indicate the timing and depth needed to produce the $\sim 200$ ppm event in the SPOC photometric aperture. We found no evidence of an NEB that might be responsible for the SPOC detection. By a process of elimination, we conclude that the transit is likely occurring on TOI-1444.

\section{Radial Velocity Analysis} \label{sec:rv}
We obtained 59 high-resolution spectra of TOI-1444 from UTC Dec 15 2019 to Dec 25 2020 on the High Resolution Echelle Spectrometer on the 10m Keck I telescope \citep[Keck/HIRES][]{HIRES}. Two of these are iodine-free spectra which serve as the template for radial velocity extraction. The rest were obtained with the iodine-cell in the light path to serve as the source for wavelength calibration and the reference for the line spread function. Each exposure of TOI-1444 was about 15 min achieving a median SNR of 140 per reduced pixel near 5500 Å. Whenever possible, we tried to obtain multiple exposures within each night; this helps to separate the radial velocity variation due to the short-period planet b from any longer-period stellar activity contamination. Such a strategy has been employed in the RV follow-up of many USP planets \citep[e.g.][]{Howard78}. The radial velocity variation was extracted using the forward-modeling Doppler pipeline described in \citet{Howard}. To quantify the stellar activity of TOI-1444, we analyzed the Ca II H\&K lines and extracted the $S_{\rm HK}$ using the method of \citet{Isaacson}.  We looked for any correlation between the measured RV and activity index $S_{\rm HK}$. We applied a Pearson correlation test which reported a $p$-values of 0.65 i.e. a lack of correlation between the measured RV and $S_{\rm HK}$. This again testifies the low activity of TOI-1444. The extracted RV and stellar activity indices are presented in Table 4.

\begin{figure*}
\begin{center}
\includegraphics[width = 1.7\columnwidth]{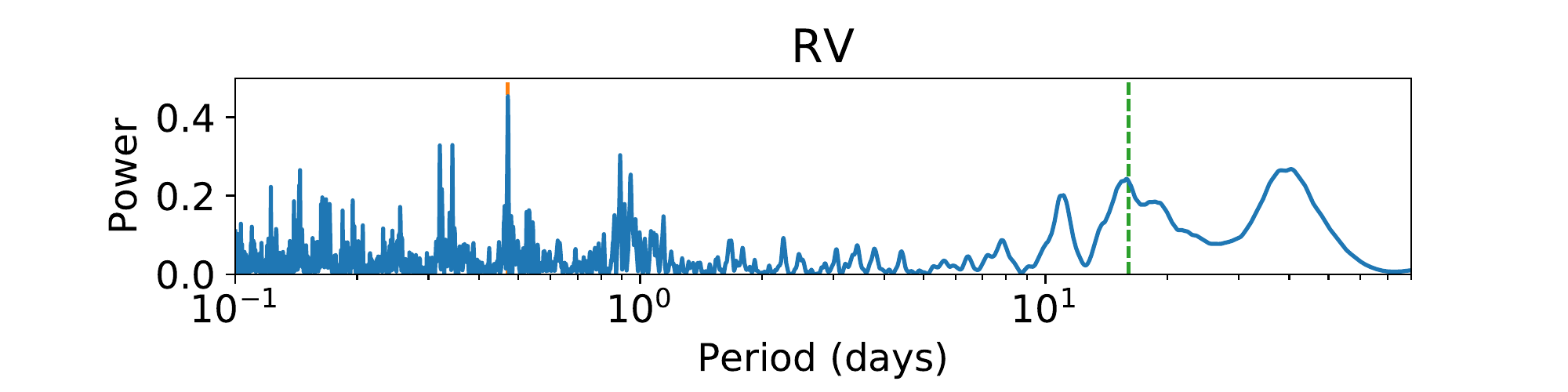}
\includegraphics[width = 1.7\columnwidth]{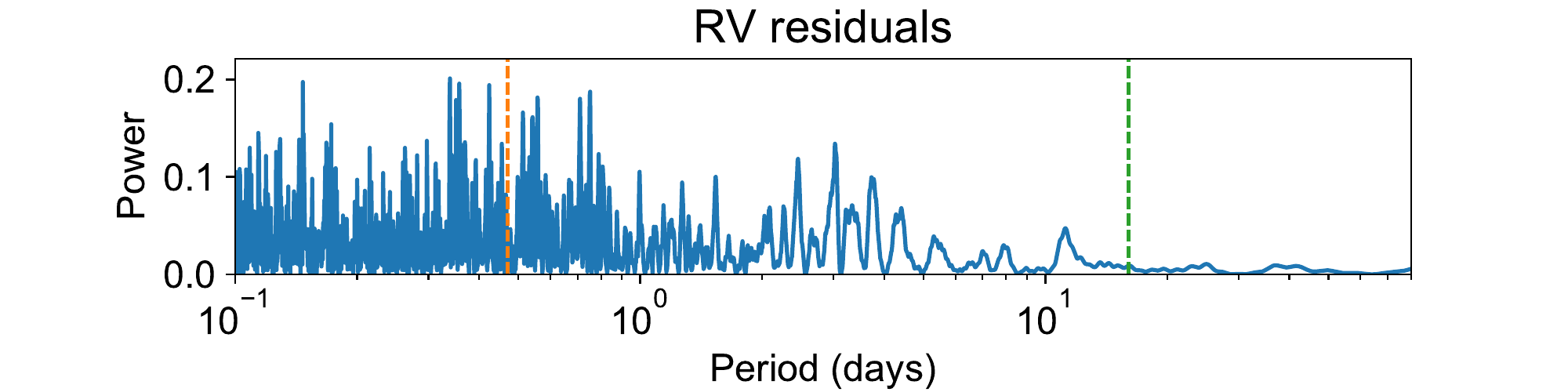}
\includegraphics[width = 1.7\columnwidth]{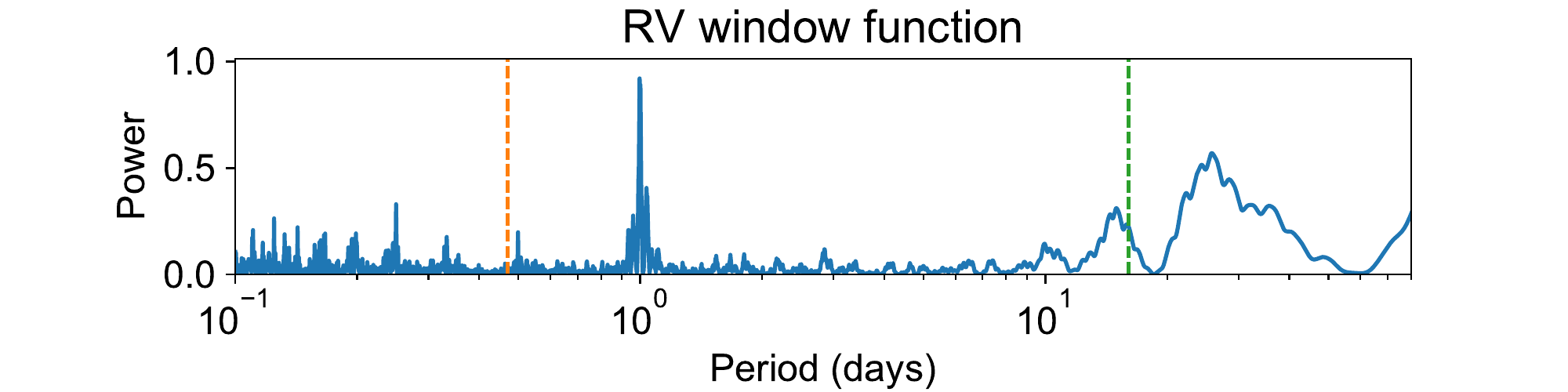}
\includegraphics[width = 1.7\columnwidth]{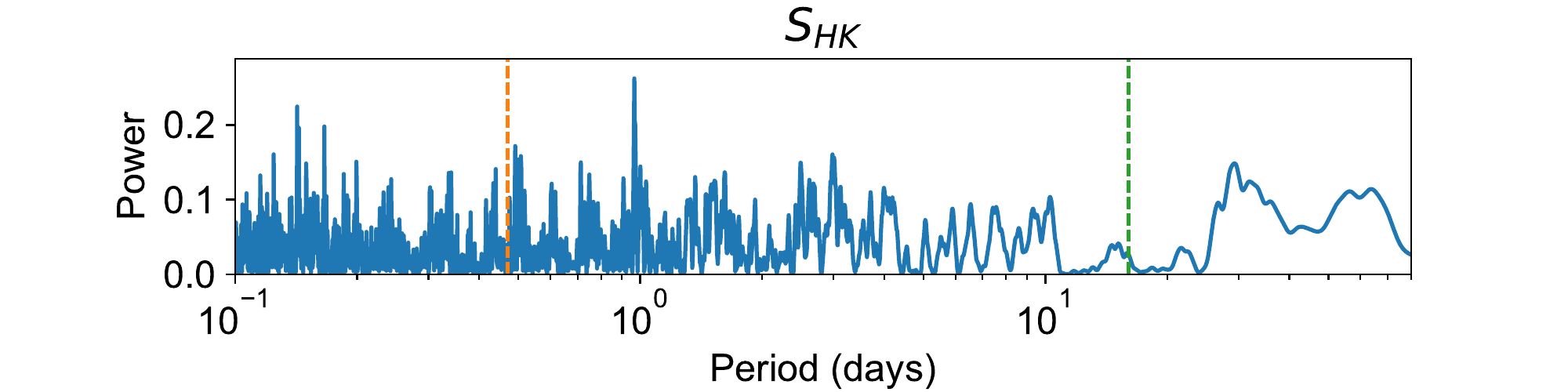}
\caption{The Lomb-Scargle periodograms of RV datasets, the RV residuals after removing the best-fit 2-planet model, the RV window function and chromospheric activity index $S_{HK}$. The vertical dotted lines show the orbital periods of the transiting planet b and the non-transiting planet candidate c. The peak due to planet c is clearly seen in the RV dataset but not in the chromospheric activity index. Another peak in the RV dataset near 38 days is likely due to an alias of the monthly variation.}
\label{fig:periodogram}
\end{center}
\end{figure*}

\begin{figure}
\begin{center}
\includegraphics[width = .9\columnwidth]{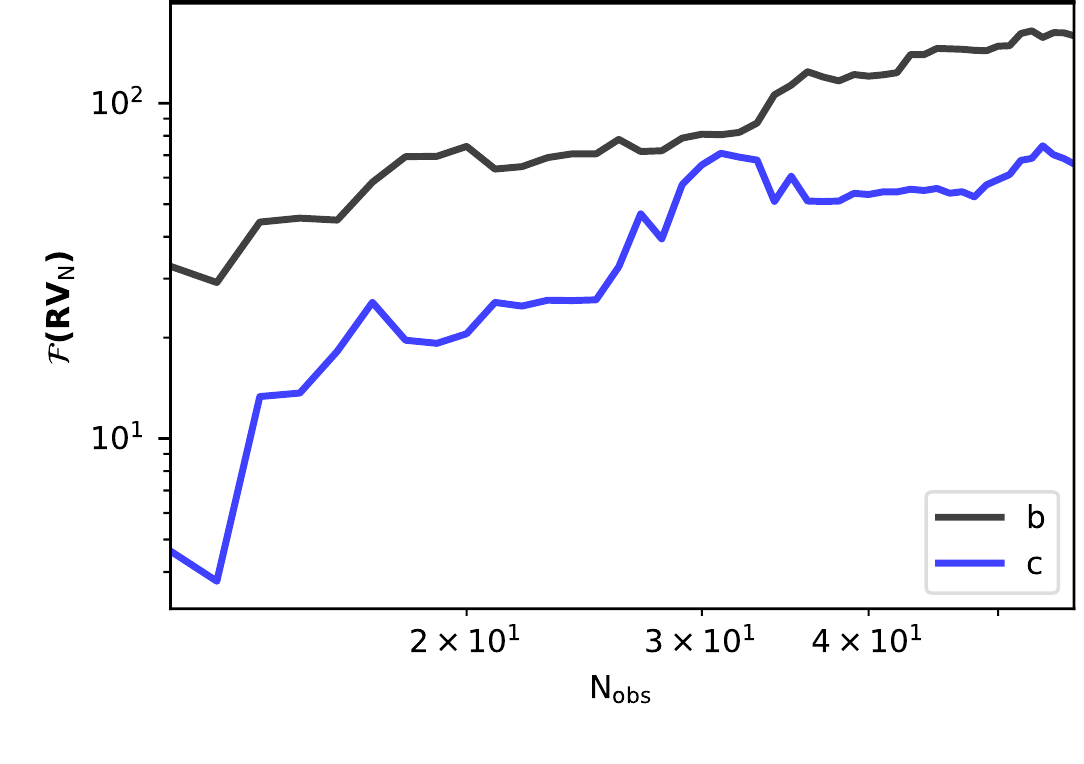}
\caption{The strength of the RV signal of planet b and c in Lomb-Scargle periodogram as a function of number of data points included. The steady increase of signal strength over data points indicates coherence over time and favors the planetary interpretation.}
\label{fig:periodogram_Nob}
\end{center}
\end{figure}

\begin{figure*}
\begin{center}
\includegraphics[width = 1.8\columnwidth]{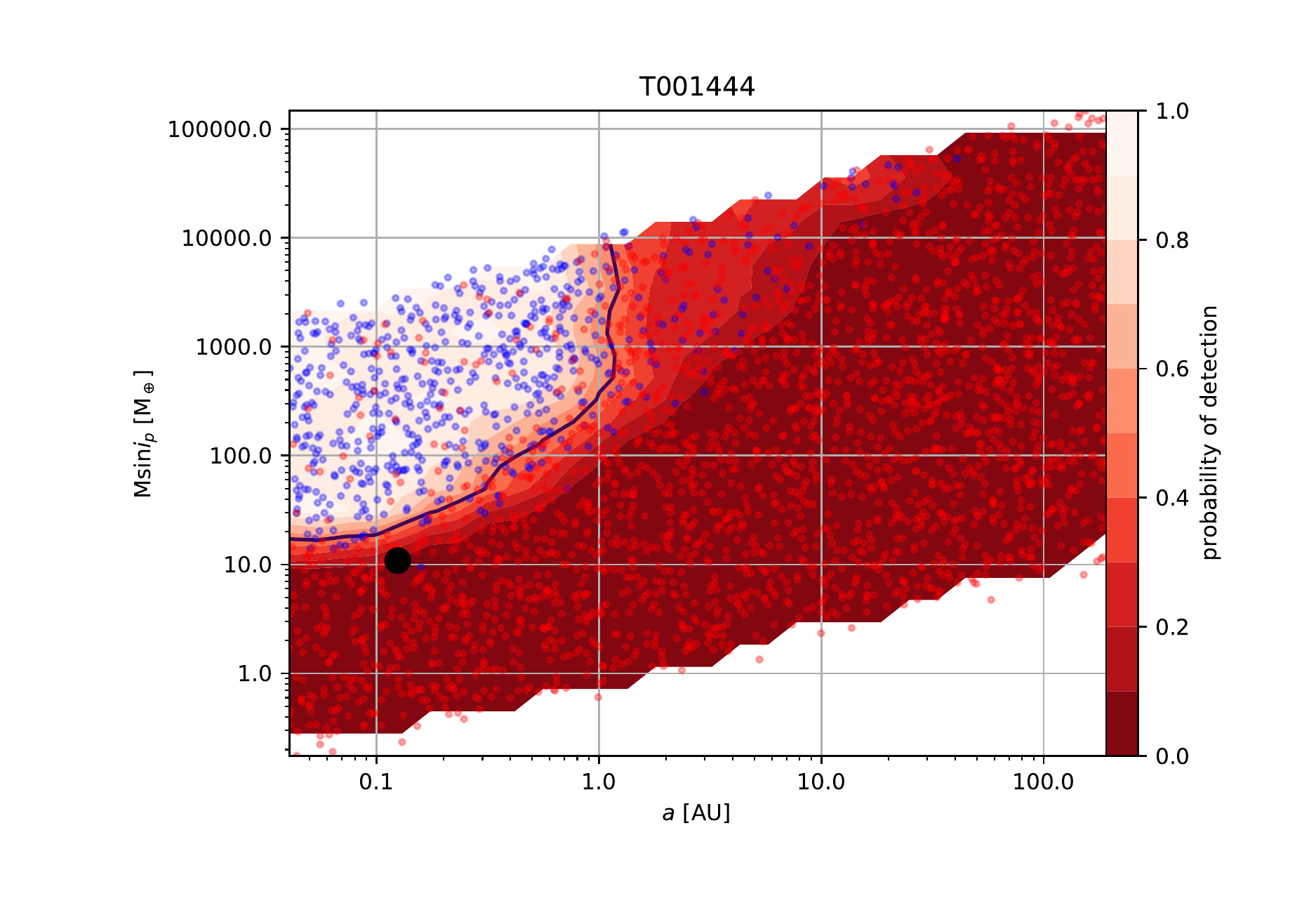}
\caption{{\tt RVsearch} completeness contour of TOI-1444 for a third planetary signal in the existing RV dataset. The circles are injected planetary signals that are recovered (red circles) or missed (blue circles) by {\tt RVsearch}. The completeness contours are based on this injection/recovery test. The black circle is the candidate planetary signal for planet c around 16 days. Given the current RV dataset, a third sub-Neptune planets within 1AU could remain hidden from detection.}
\label{fig:completeness}
\end{center}
\end{figure*}

\subsection{Non-transiting planet c?}
We first applied the {\tt Python} package {\tt RVSearch} \footnote{https://github.com/California-Planet-Search/rvsearch} to determine the number of planetary signals in our RV dataset. In short, {\tt RVSearch} sequentially look for peaks in the LS periodogram of the RV dataset after removing the best-fit Keplerian model of planetary signals detected in previous iterations. The code stops when the Bayesian Information Criterion ($\Delta$ BIC) no longer favors the addition of another planetary signal. {\tt RVSearch} has been widely tested on known planetary systems. For more detail on {\tt RVSearch}, we refer interested readers to Rosenthal et al (2021).

Applying {\tt RVSearch} to TOI-1444, a strong peak at the transiting period of TOI-1444b is recovered (Fig.\ref{fig:periodogram}). {\tt RVSearch} detects a candidate 16-day signal. No corresponding peak is seen in the RV window function or the activity index $S_{HK}$. Moreover, the strength of the 16-day periodicity steadily increased as we included more and more RV data in the periodogram. In contrast, a signal caused by stellar activity loses coherence over time because the underlying surface magnetic activity typically emerge and subside on weeks-to-months timescale for solar-like stars.

{\tt RVSearch} can also perform injection and recovery test of planetary signals in the $M_p$sin$i$-$P_{\rm orb}$ plane. The result is a completeness contour showing the sensitivity of the RV dataset at hand to planetary signals of different strength and periodicity in addtion to the detected planets (e.g. Fig. \ref{fig:completeness}). Given the current HIRES RV dataset for TOI-1444, we were unable to identify a third planetary signal. The completeness contour for that third planet is shown in Fig. \ref{fig:completeness}. With the current HIRES dataset, other sub-Neptune planets (<10$M_\oplus$) within 1AU of TOI-1444 can easily remain hidden from detection.

\subsection{{\tt RadVel} Model}
We first modeled the RV dataset using Keplerian models. We used the publicly available {\tt Python} package {\tt RadVel} \citep{RadVel}. Each planetary signal is described by its orbital period $P_{\rm orb}$, time of inferior conjunction $T_c$, eccentricity $e$, argument of periastron $\omega$ and the RV semi-amplitude $K$. We allowed an RV offset $\gamma$, a linear RV trend $\Dot{\gamma}$ and we included a jitter term to encapsulate any additional astrophysical or instrumental noise. We reparameterized $e$ and $\omega$ into $\sqrt{e}$cos$\omega$, $\sqrt{e}$sin$\omega$. Since the ephemeris is much better constrained with the transit data, we imposed Gaussian priors on $P_{\rm orb}$ and  $T_c$ using the results from Section \ref{sec:transit}. We imposed uniform priors on the RV semi-amplitude $K$, the jitter  $\sigma_{\text{jit}}$, $\sqrt{e}$cos$\omega$ (with range [-1,1]), $\sqrt{e}$sin$\omega$ ([-1,1]), $\gamma$ and $\Dot\gamma$. Our likelihood function is as follows:

\begin{equation}\label{rv_likelihood}
\mathcal{L}=  \prod_{i}\left({\frac{1}{\sqrt{2 \pi (\sigma_i^2 + \sigma_{\text{jit}}^2)}}}  \exp \left[ - \frac{[RV(t_i) - \mathcal{M}(t_i)]^2}{2 (\sigma_i^2 + \sigma_{\text{jit}}^2)} \right] \right)
\end{equation}

where $RV(t_i)$ is measured RV; $\mathcal{M}(t_i)$ is the sum of Keplerian planetary signals, a constant RV offset $\gamma$ and a linear RV trend $\Dot{\gamma}t_i$; $\sigma_{i}$ is the internal uncertainty.

We varied the number of planets in our model; and we tested if the current dataset supports non-zero eccentricity and $\Dot{\gamma}$. We selected the best model using both Bayesian Information Criterion ($\Delta$ BIC) and Akaike information criterion ($\Delta$ AIC). Our best-fit model (lowest $BIC$ and $AIC$) contains the RV signals of planet b and a non-transiting planet with a 16-day orbit identified by {\tt RVSearch}. Non-zero eccentricity for either planet was not preferred by the current dataset. A linear RV trend of $\Dot{\gamma}=-0.008\pm0.012$m~s$^{-1}$ is marginally disfavored by the current dataset with $\Delta BIC\approx-2$. We then performed an MCMC analysis to sample to posterior distribution of the various parameters. The sampling procedure is similar to that descrbied in Section \ref{sec:transit}. We summarize the posterior distribution in Table 2.

\begin{figure*}
\begin{center}
\includegraphics[width = 1.5\columnwidth]{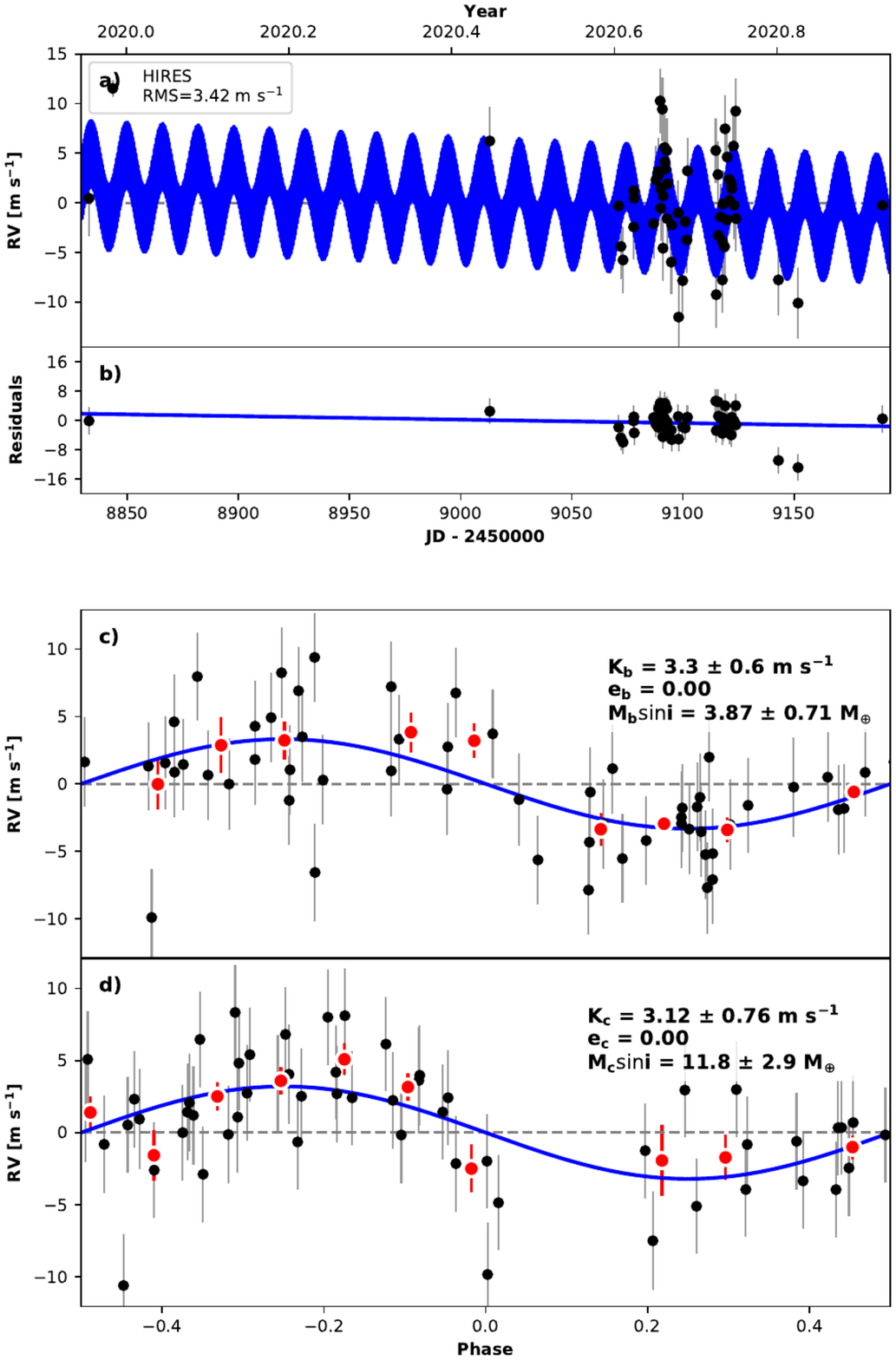}
\caption{The best-fit {\tt RadVel} 2-planet model of TOI-1444. The upper panel shows the RV data point over time versus the combined 2-planet model. The lower two panels show the phase folded RV variation after removing the signal of the other planet. The two planets have orbital periods of 0.47 and 16 days.}
\label{fig:rv_RadVel}
\end{center}
\end{figure*}

\subsection{Gaussian Process Model}

The RV residuals after removing the best 2-planet Keplerian model still shows a root-mean-square variation (RMS) of 3.4 m~s$^{-1}$ which is substantially larger than the internal uncertainties in Table 4. Moreover, the  RV residuals visually display some correlated noise component in Fig. \ref{fig:rv_RadVel}. We investigated if these residual noises can be tidied up by a Gaussian Process (GP) model \citep[e.g. ][]{Haywood,Grunblatt2015} and and disentangle the planetary signal from the stellar activityf.

We adopted the GP model used in our previous works \citep{DaiK2,Dai2019}. Stellar surface magnetic activity rotating in and out of view of the observer is chiefly responsible for the correlated stellar noise in RV measurements, the quasi-sinusoidal flux variation in the light curve and the variation of chromospheric activity index $S_{HK}$ \citep[e.g.,][]{Aigrain2012}. In other words, these effects stem from the same underlying physical process, we model them with a common GP model. The plan of attack is to train a quasi-periodic GP model on the out-of-transit {\it TESS} light curves and the HIRES $S_{HK}$ index, before applying it the RV dataset. Our kernel has the following form:
\begin{equation}
\label{covar}
C_{i,j} = h^2 \exp{\left[-\frac{(t_i-t_j)^2}{2\tau^2}-\Gamma \sin^2{\frac{\pi(t_i-t_j)}{T}}\right]}+\left[\sigma_i^2+\sigma_{\text{jit}}^2\right]\delta_{i,j}
\end{equation}

where $C_{i,j}$ represents the covariance matrix; $\delta_{i,j}$ is the Kronecker delta function;
$h$ is the covariance amplitude; $t_i$ is the time of $i$th observation; $\tau$ is the correlation timescale; $\Gamma$ is the relative importance between the squared exponential and periodic parts of the kernel; and $T$ is the period of the covariance; $\sigma_i$ is the reported uncertainty and $\sigma_{jit}$ is a jitter term. Our likelihood function is as follows:

\begin{equation}
\label{likelihood}
\log{\mathcal{L}} =  -\frac{N}{2}\log{2\pi}-\frac{1}{2}\log{|\bf{C}|}-\frac{1}{2}\bf{r}^{\text{T}}\bf{C} ^{-\text{1}} \bf{r}
\end{equation}
where $\mathcal{L}$ is the likelihood function; $N$ is total number of RV data points; $\bf{C}$ represents the covariance matrix; and $\bf{r}$ is the residual the observed RV minus the model RV.

We ran an MCMC analysis (similar to that described in Section \ref{sec:transit}) on the {\it TESS} light curves to constrain the posterior distribution of the various hyperparameters. As we mentioned earlier, we could not robustly detect the stellar rotation period in the {\it TESS} light curve from a periodogram analysis. This is may be due to the low activity of the host star. Correspondingly, our GP model of the {\it TESS} photometry yielded very broad posterior distribution on various hyperparameters. We then tried if the addition of HIRES $S_{HK}$ index gave better constraints on the hyperparameters. $S_{HK}$ was sparsely sampled and only varied very mildly with RMS of about 0.004. Consequently, $S_{HK}$ did not significantly improve the constraints on the GP hyperparameters. 

\begin{figure*}
\begin{center}
\includegraphics[width = 1.5\columnwidth]{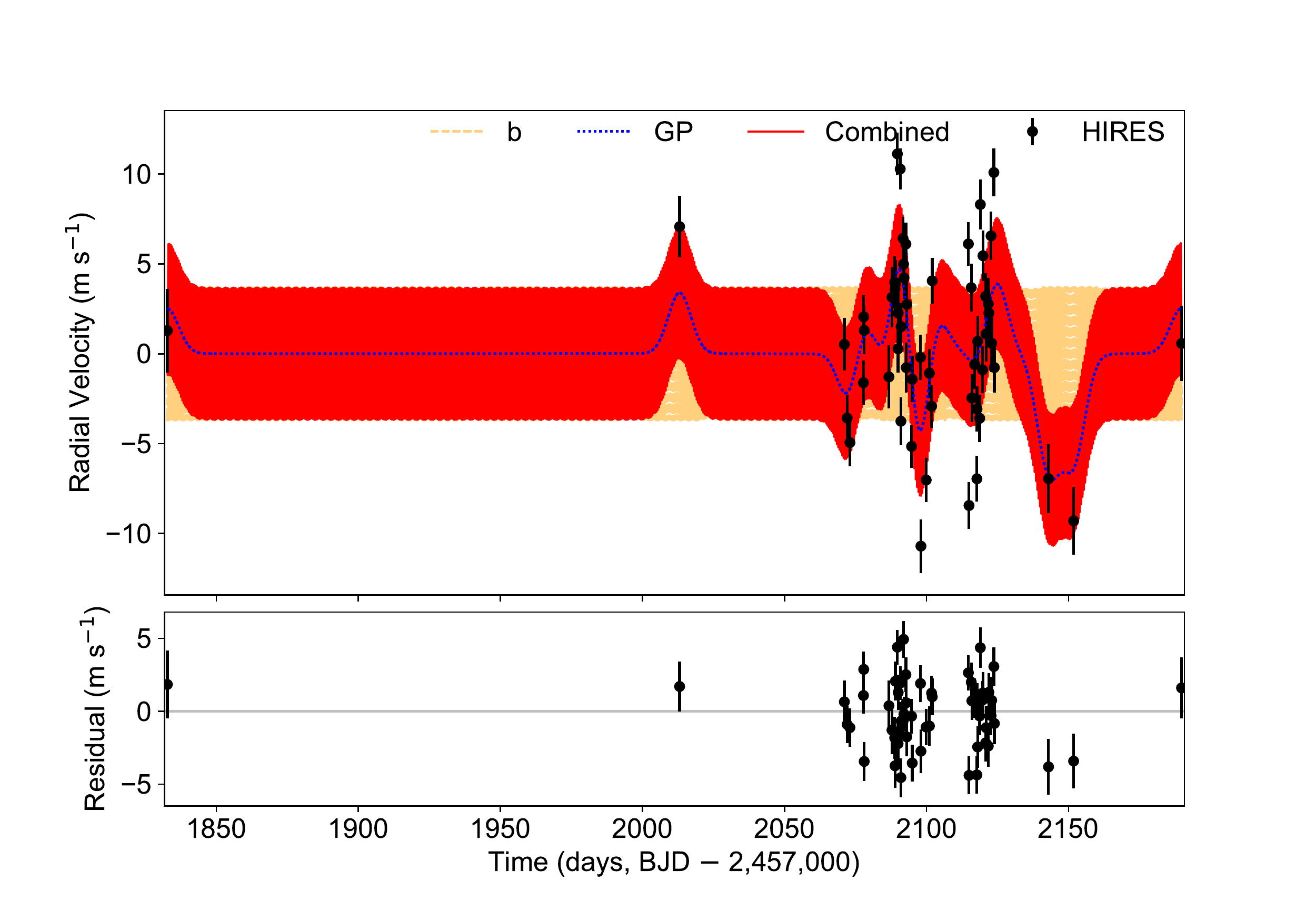}
\includegraphics[width = 1.5\columnwidth]{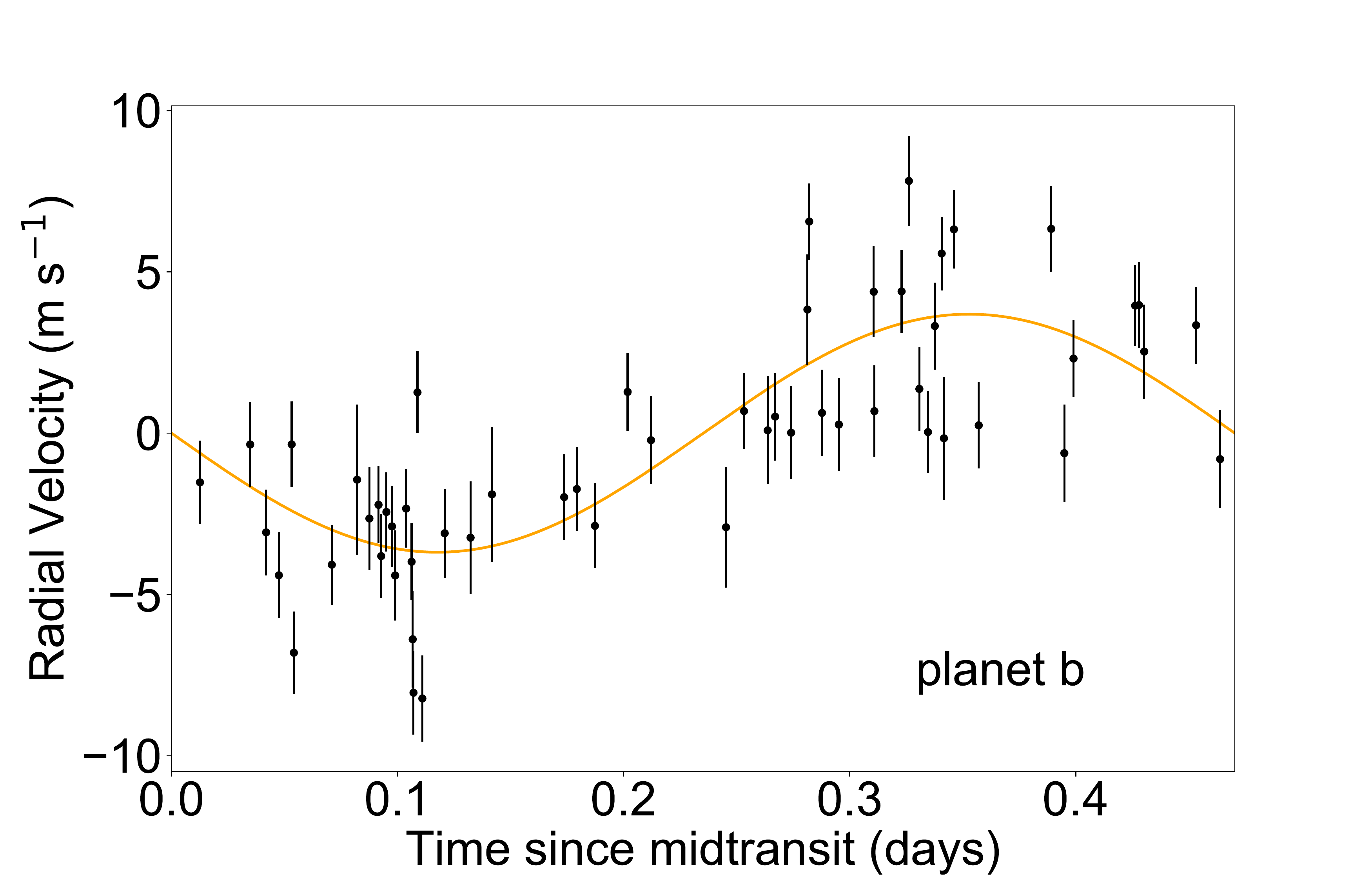}
\caption{The Gaussian Process Model of TOI-1444. Only planet b is included in this model because stellar rotation period was not securely detected in the light curve. Without a prior on stellar rotation period, the GP model is so flexible that it tends subsume the signal from planet c. Nonetheless, planet b is securely detected in the GP model with mass constraint consistent with the {\tt RadVel} model.}
\label{fig:rv_gp}
\end{center}
\end{figure*}

Therefore, we chose to let the hyperparameters float freely in our final GP analysis of the RV dataset. The exception is that we did limit $T$  the covariance period (a proxy for stellar rotation period) between 1 and 200 days to avoid inference with the 0.47-day planet b. Understandably, without an imposed prior on the various hyperparameters, the flexibility of this GP model subsumed the signal of the candidate 16-day planet c. Planet c only has an upper limit $K_c<4.3$m~s$^{-1}$ at 95\% confidence level in our GP model. In fact, the lowest BIC GP model only included the signal from planet b (Fig. \ref{fig:rv_gp}). An MCMC analysis showed that the posterior distribution of the semi-amplitude of planet b is $K_b = 3.59 \pm 0.75$m~s$^{-1}$ which agrees with the $K_b = 3.30^{+0.58}_{-0.59}$m~s$^{-1}$ found by {\tt RadVel}. Since the GP model has far more complexity than the {\tt RadVel} model and the $K$ value came out less well constrained, we adopted the results from {\tt RadVel} for further analyses in this work.

\begin{deluxetable*}{lll}
\tablecaption{Planetary Parameters of TOI-1444} \label{tab:planet_para}
\tablehead{
\colhead{Parameter}  & \colhead{Symbol} &  \colhead{Posterior Distribution} }
\startdata
Planet b \\
Planet/Star Radius Ratio & $R_p/R_\star$  & $0.01410^{+0.00060}_{-0.00058}$  \\
Time of Conjunction (BJD-2457000) & $t_c$  & $1711.3676\pm0.0016$  \\
Impact Parameter & $b$  & $0.38\pm0.14$ \\
Scaled Semi-major Axis & $a/R_\star$  & $2.73^{+0.078}_{-0.074}$  \\
Orbital Inclination (deg) & $i$  & $82^{+3}_{-2}$  \\
Orbital Eccentricity  & $e$  & 0 (fixed)  \\
Orbital Period (days) & $P_{\rm orb}$   & $0.4702694\pm0.0000044$  \\
Semi-amplitude (m/s)  & $K$ &$3.30^{+0.58}_{-0.59}$ \\
Planetary Radius ($R_\oplus$)  & $R_{\rm p}$ &1.397$\pm0.064$ \\
Planetary Mass ($M_\oplus$)  & $M_{\rm p}$ &3.87$\pm0.71$ \\
Secondary Eclipse Depth (ppm)& $\delta_{\rm sec}$ & $27\pm12$\\
Time of Secondary Eclipse (days)& $t_{\rm sec}$ & $0.236\pm0.019$\\
Amplitude of Illumination Effect (ppm)& $A_{\rm ill}$ & $24\pm10$\\
Phase Offset of Illumination Effect ($^\circ$)& $\theta_{\rm ill}$ & $16\pm27$\\
\hline
Planet c \\
Time of Conjunction (BJD-2457000) & $t_c$  & $ 713.0\pm2.0$  \\
Orbital Period (days) & $P_{\rm orb}$   & $16.066 \pm 0.024$  \\
Orbital Eccentricity  & $e$  & 0 (fixed)  \\
Semi-amplitude (m~s$^{-1}$)  & $K$ &$3.12^{+0.75}_{-0.76}$ \\
Projected Planetary Mass ($M_\oplus$)  & $M_{\rm p}\sin i$ &11.8$\pm$2.9 \\
\hline
RV Jitter (m~s$^{-1}$)  & $\sigma$ &$3.2 \pm 0.4$\\
\enddata
\end{deluxetable*}

\section{Discussion}

\subsection{Are the known USPs still predominantly rocky?}
\begin{figure}
\begin{center}
\includegraphics[width = 1.\columnwidth]{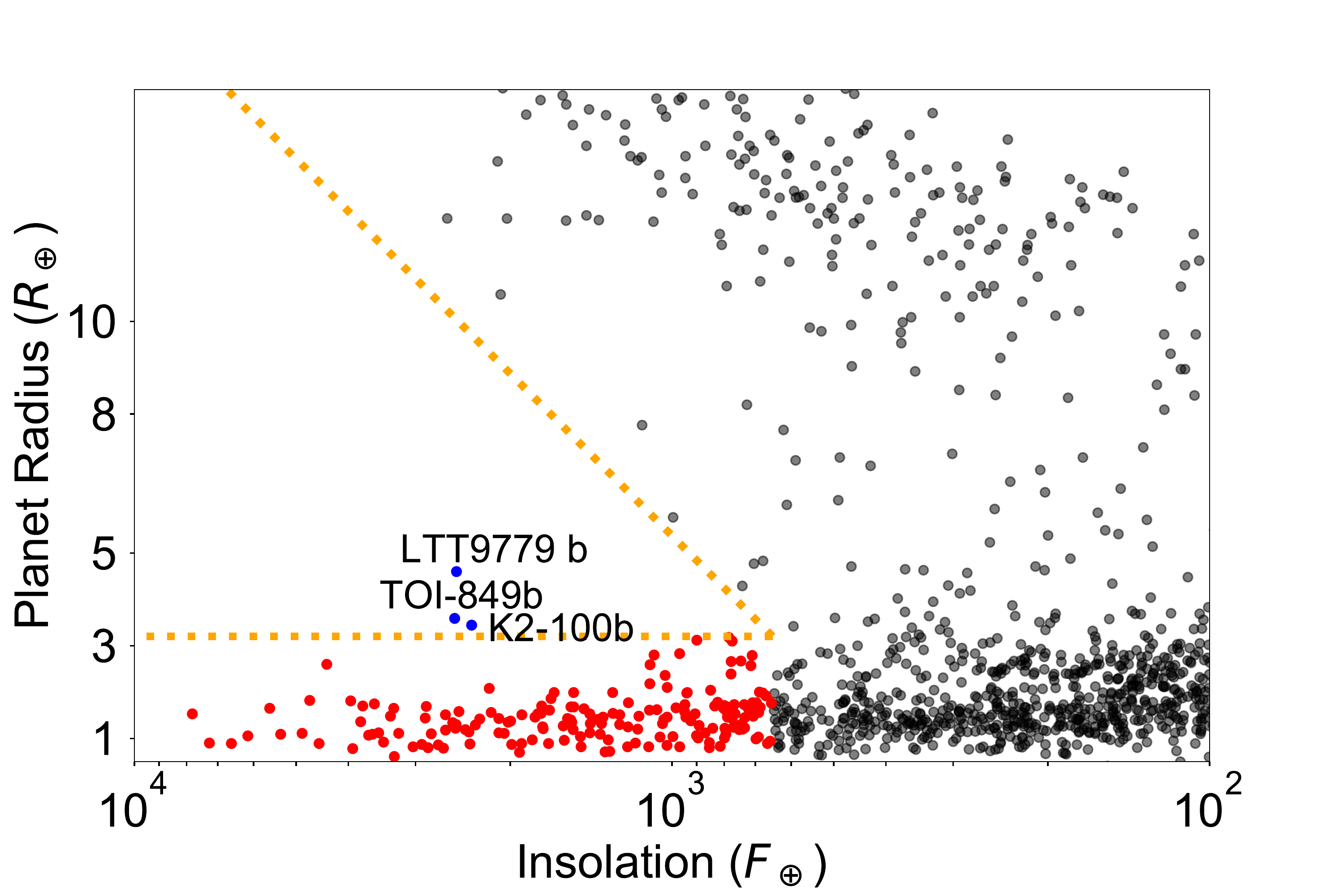}
\caption{The planetary radius and insolation of all confirmed transiting exoplanets. The orange dotted lines empirically outline the so-called ``hot Neptune Desert'' above 650$F_\oplus$\citep{Lundkvist}. There are three hot Neptunes (blue points) that apparently straggle this desert which we contrast with the smaller planets that have presumably lost its H/He envelope (red points).}
\label{fig:insolation}
\end{center}
\end{figure}

Before any discussion, we would like to loosely define USPs as terrestrial planets ($<2R_\oplus$) with orbital period less than one day (the prevailing definition in the literature) as well as those with an insolation level larger than 650$F_\oplus$ (see Table 3 for the complete list).  650$F_\oplus$ is an empirical boundary, identified by \citet{Lundkvist,USP}, beyond which photoevaporation is so strong that any Neptune-sized planets are quickly stripped down to their rocky cores by the strong stellar irradiation (Fig. \ref{fig:insolation}), thereby creating a ``Hot Neptune Desert''.

\citet{Dai2019} performed a uniform analysis of all transiting USPs that also have RV mass constraints. In particular, they used Gaia parallax information to better constrain the host stellar properties. The increased precision on stellar radius translated to increased precision on planetary radii. Moreover, the mean stellar density from Gaia and spectroscopy further disentangled degeneracies in transit modeling. The results significantly improved constraints on planetary radii. For example, the radius constraint of Kepler-78 b improved from $R_p = 1.20\pm0.09 R_\oplus$ in \citet{Howard2013} to $R_p = 1.228^{+0.018}_{-0.019} R_\oplus$ with Gaia. \citet{Dai2019} also applied a Gaussian Process model uniformly to all USP planets that required mitigation of stellar activity contamination in RV datasets. The improved mass and radius constraints on the sample of 10 USPs revealed a prevalence of 35\%Fe-65\%MgSiO$_3$ Earth-like composition. 

\begin{figure*}
\begin{center}
\includegraphics[width = 1.5\columnwidth]{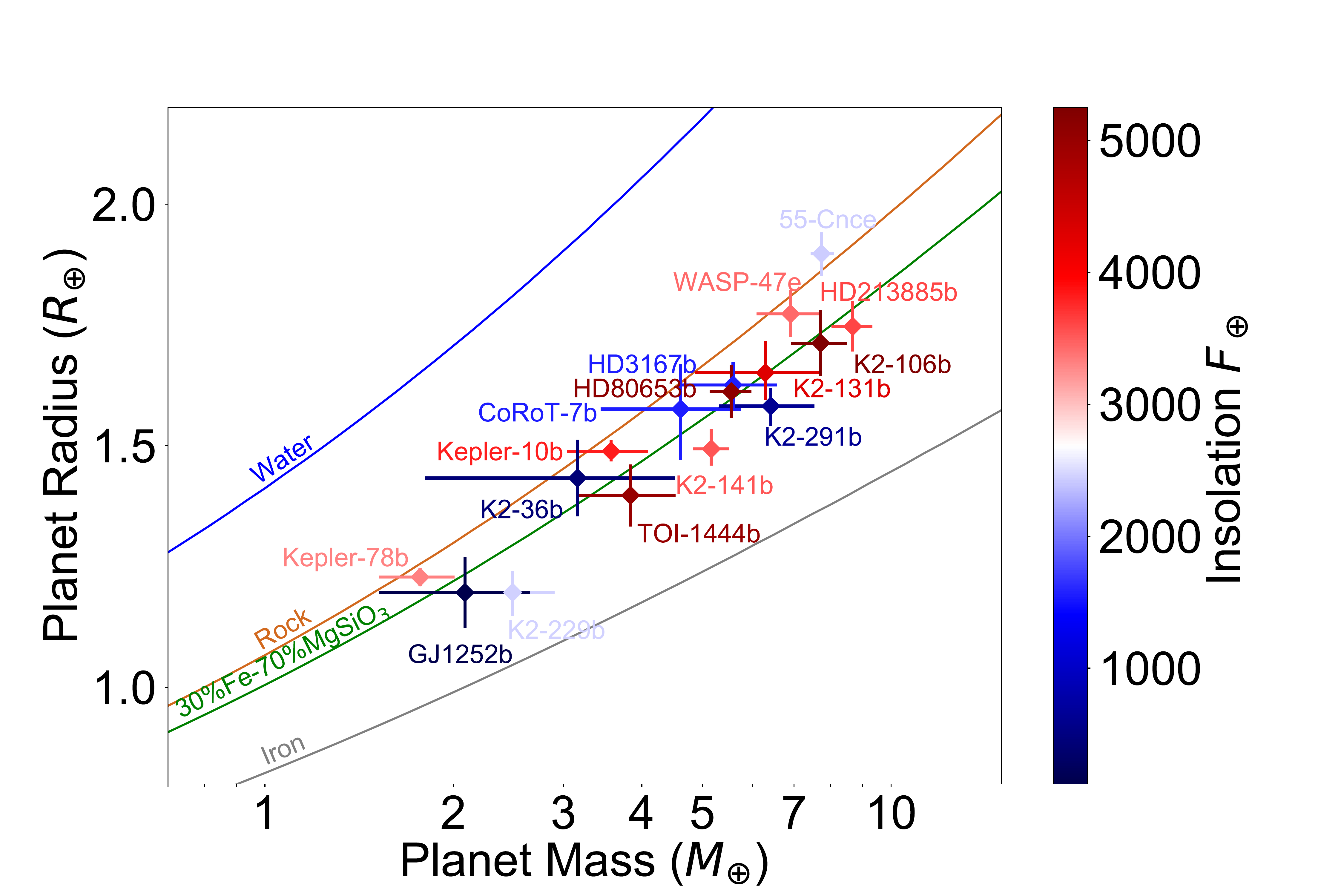}
\includegraphics[width = 1.5\columnwidth]{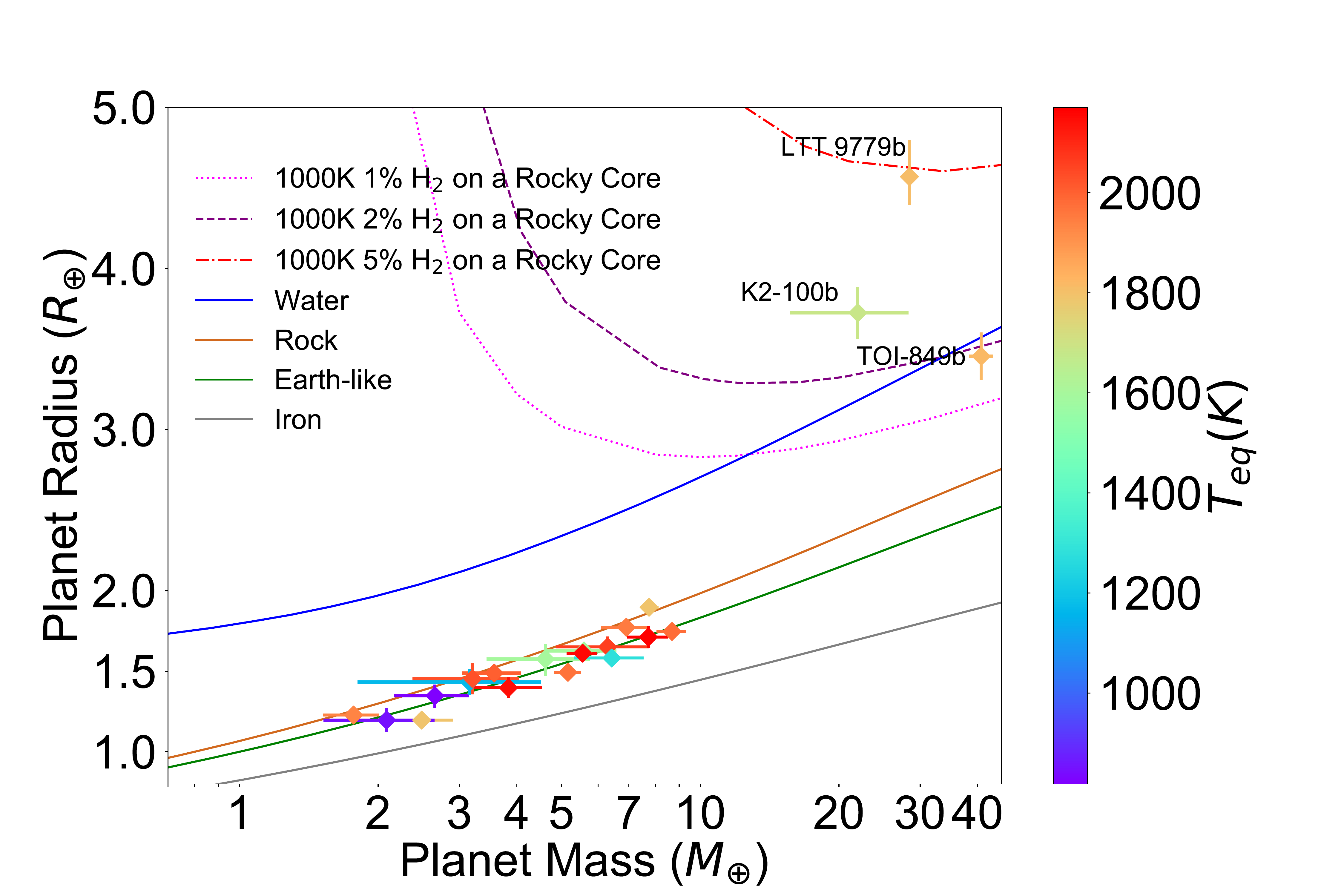}
\caption{The measured mass and radius of all USP planets along with theoretical mass-radius relation from \citet{Zeng2016}. The lower panel includes the three ultra-hot Neptunes as well. The USP cluster around an Earth-like composition of iron-rock mixture, whereas hot Neptunes require H/He atmosphere or volatile envelope to explain their measured mass and radius.}
\label{fig:mass_radius}
\end{center}
\end{figure*}

In this work, we applied the same set of analysis to TOI-1444b and interpret the resultant mass and radius constraint along with other USP planets. The interest in USP planets has boomed in recent years, the number of USP RV mass measurements increased from 10 in \citet{Dai2019} to 17 at the time of writing of this work. Moreover, the inclusion of Gaia parallax information and Gaussian Process modeling has become a standard practice in these new USP papers (Table 3). This puts the new USPs reported by different groups on relatively equal footing and ready for comparison. In Fig. \ref{fig:mass_radius} upper panel, we show the mass and radius of all USP planets with RV mass measurements. At a glance, USPs seem to cluster around an Earth-like composition. To quantify the compositions, we adopted a simple two-layer
model where planets have an iron core and a MgSiO$_3$ mantle. We used the procedure described by \citet{Zeng2016} to convert the measured mass and radius of the planet to a Fe-MgSiO$_3$ ratio. For an individual planet, the confidence interval is relatively wide e.g. TOI-1444b can have 43$\pm$30\% of its mass in iron given our mass and radius measurement. However, as an ensemble, the USP planets generally cluster around the Earth-like composition with a weighted mean of 32$\pm4\%$ mass in an iron core. This is consistent with the general picture that planets at the lower peak of the bimodal radius distribution are predominantly rocky \citep{Rogers,Dressing,Fulton}.

We note that we have excluded two USPs from the averaging of iron core mass fraction. The mass of TOI-561 reported by \citet{Weiss561} and \citet{Lacedelli} differ by almost a factor of two.  More RV data are being taken to resolve this discrepancy (Brinkman et al, in prep). By focusing on the USP planets i.e. planets that are most strongly irradidated, we hope to probe the exposed planetary cores directly without worrying about the degeneracy introduced by a planetary atmosphere. This assumption has held up well for most USPs in our sample. We examined the composition of USP against the insolation level. If a substantial atmosphere were present on an USP, one may expect the scale height to vary strongly with insolation level. The scale height variation would have translated to a correlation between insolation level and the inferred planetary composition. However no correlation between insolation and iron core mass fraction was found (Fig. \ref{fig:mass_radius}). That said, we did exclude 55 Cnc e, one of the largest and coolest USPs, from our analysis. The strong phase offset seen in Spitzer observation of 55 Cnc e \citep{Demory} demands a strong heat circulation between the day and the night side of the planet that, as \citet{Angelo} argue, requires the presence of an atmosphere on 55 Cnc e. The larger planet mass and the cooler equilibrium temperature of 55 Cnc e may help retain its atmosphere compared to other USPs in the current sample.

\subsection{USP versus Ultra-hot Neptunes}\label{Sec:usp_vs_nep}
What is the relation between the USP planets and the recently reported ultra-hot Neptunes K2-100 b \citep{Mann100}, LTT 9779 b \citep{Jenkins}, TOI-849 b\citep{Armstrong849}? Are they the same group of planets that only differ in size? In this section, we will show that USPs and ultra-hot Neptunes differ in planetary/stellar properties and system architecture suggesting that they are probably two distinct groups of planets.

We identified a sample of three very hot Neptunes between 3-5 $R_\oplus$ and with an insolation level $>2000F_\oplus$ (Fig. \ref{fig:insolation}). Again $>650F_\oplus$ is the empirical boundary of ``Hot Neptune Desert'' proposed by \citet{Lundkvist}. $>2000F_\oplus$ securely put these planets within the ``Desert''.  Currently, there are three confirmed hot Neptunes straggling in this so-called "Hot Neptune Desert": LTT 9779 b\citep{Jenkins}, TOI-849 b\citep{Armstrong849}, and K2-100b \citep{Mann100} that have well-measured mass and radius. We put these hot Neptunes in the same mass radius diagram with the USP planets (Fig. \ref{fig:mass_radius} lower panel). Clearly, the USPs and the hot Neptunes occupy different part of the mass-radius parameter space. The USPs are all below 10$M_\oplus$, even though RV mass measurements bias towards the detection of heavier planets (a point we will return to later). The USPs also cluster around an Earth-like composition of iron-rock mixture. However, the three close-in hot Neptunes are near or above 100\% water composition line in the mass radius diagram. This indicates the presence of a substantial H/He atmosphere or a water envelope for these planets. If we assume that they have Earth-like rocky cores, a  1-5\% mass fraction H/He envelope is needed to reproduce the observed mass and radius. Recent Spitzer phase curve and secondary eclipse observations of LTT 9779 b pointed to the presence of a heavy-molecular weight (e.g. CO) atmosphere \citep{Crossfield9779,Dragomir9779}.

\citet{Winn2017} showed that the metallicity of USP host stars are solar-like which is similar to the longer-period sub-Neptune planets commonly discovered by {\it Kepler} (blue histogram in Fig. \ref{fig:metallicity}). On the other hand, hot Jupiters are known to preferentially occur around metal-rich systems (orange histogram in Fig. \ref{fig:metallicity}). Interestingly, the three hottest Neptunes all have similarly metal-rich host stars: LTT 9779 with [Fe/H] = 0.27$\pm0.03$ \citep{Jenkins}; K2-100 with [Fe/H] = 0.22$\pm0.09$ \citep{Mann100}; TOI-849 with [Fe/H] = 0.19$\pm0.03$ \citep{Armstrong849}. Even though the sample size is only 3, the distribution of hot Neptunes is beginning to show a stark distinction from the solar-like distribution of USP, and bears more resemblance to the metal-rich environments of hot Jupiters. This result is reminiscent of the previous work by \citet{Dong} who showed an even more statistically robust preference for hot Neptunes to occur around metal-rich host stars if we consider orbital periods as long as 10 days.

Another feature that differentiates USPs from hot Neptunes is the presence of additional planetary companions. 14 out of the 17 well-characterized USPs have other planetary companions in the same system discovered by transits or radial velocity follow-up (Table 3). We further note that the RV datasets of the three systems where the USPs remain the only detected planet (K2-131, Kepler-78 and K2-291) are heavily contaminated by activity-induced RV variation that amounts to tens of m~s$^{-1}$. We applied {\tt RVSearch} to these systems to constrain the sensitivity of the existing RV dataset to additional planetary signals (similar to Fig. \ref{fig:completeness}). {\tt RVSearch} showed that the activity-induced RV variation could easily inhibit the detection of injected planetary signals  with <$15M_\oplus$  within 1 AU. Turning our attention to the hot Neptunes, among the three hottest Neptunes, none of them have detected additional planets even though some of RV monitoring baselines extended more than 100 days (Table 3). This hints that hot Neptunes tend to be ``lonely'' similar to the hot Jupiters.

To sum up, USPs planets tend to be rocky in composition while hot Neptunes possess substantial H/He atmospheres or other volatile envelopes. USPs occur in stars with solar-like abundances, while hot Neptunes are preferentially found in metal-rich systems. USPs tend to have other close-in planetary companions while hot Neptunes tend to be lonely. These properties distinguish the two groups. Moreover, the properties of the hottest Neptunes are quite similar to their bigger cousins hot Jupiters. This similarity actually extends to many Neptune-sized planets between 2-6 $R_\oplus$ and orbital periods of  1-10 days as pointed out by \citet{Dong}. Finally, we close the section by noting that USPs, hot Jupiters and <10-day hot Neptunes all occur at about 1\% level around sun-like stars \citep[e.g.][]{USP,Cumming,Dong}. However, the hot Neptunes on sub-day orbits seem rarer. The {\it Kepler} USPs from \citet{USP} represent a uniformly studied sample amenable to occurrence rate studies. Among the 67 confirmed/validated USP candidates only 1 (KOI-3913) has a radius in the Neptune regime >3$R_\oplus$. The rest are all below 2$R_\oplus$ (Fig. \ref{fig:metallicity}). Even though this is still small number statistics, the {\it Kepler} USP sample suggests that hot Neptunes on sub-day orbit are probably as rare as 0.1-0.01\% level around sun-like star. Future {\it TESS} occurrence rate studies will firm up this number. Another system worth mentioning is NGTS-4b \citep{West} a 3.2$R_\oplus$ hot Neptune on a 1.3-day orbit around a K star. It is less strongly irradiated compared to K2-100b, TOI-849 b and LTT 9779 b, right on the edge of the ``hot Neptune Desert''. It shows many similarities with K2-100b, TOI-849 b and LTT 9779 b: likely enshrouded by an atmosphere, having no other planetary companions. However, the metallicity of its host star was reported to be surprisingly low: reported in units of total metal [M/H] = -0.28$\pm0.10$. We encourage future observation to confirm the reported low metallicity.

\begin{deluxetable*}{lcp{8cm}l}
\label{tab:usp}
\tablecaption{Additional planetary companions in USP and hot-Neptune systems ranked by planet mass}
\tablehead{
\colhead{System} & \colhead{Additional Planets Detected?} & \colhead{Comments} & \colhead{Reference} }
\startdata
Kepler-78&	N	&	Activity-induced RV variation (50 m~s$^{-1}$ peak to peak, 12.5-day periodicity) prevents detection of additional planets & \citet{Howard78}\\
GJ 1252&	Y&	15 m/s drift over 10 days likely due to an additional planet 	&\citet{Shporer}\\
K2-229&	Y &	2 additional transiting planets on 8 and 31-day orbit 	&\citet{2018NatAs...2..393S}\\
LTT 3780&	Y	&	1 additional transiting planet on 12-day orbit &\citet{Cloutier}\\
K2-36&	Y	&	1 additional transiting planet on 5-day orbit &\citet{Damasso}\\
Kepler-10&	Y&	1 additional transiting planet on 45-day orbit 	&\citet{Dumusque}\\
TOI-1444&	Y&	1 additional non-transiting planet on 16-day orbit 	&This work\\
CoRoT-7&	Y	&	1 additional non-transiting planet on 3.7-day orbit &\citet{Haywood}\\
K2-141&	Y	&	1 additional non-transiting planet on 7-day orbit &\citet{Malavolta}\\
HD 3167&	Y	&	1 additional non-transiting planet on 8.5-day orbit and 1 transiting planet on 29-day orbit &\citet{Christiansen2017}\\
HD 80653&	Y	&	RV drift suggests one additional planet of 0.37$\pm0.08$ $M_{\rm Jup}$(a$/$AU)$^{-2}$  &\citet{Frustagli}\\
K2-131&	N	&	Activity-induced RV variation (60 m~s$^{-1}$ peak to peak, 3-day periodicity) prevents detection of additional planets &\citet{DaiK2}\\
K2-291&	N	&	Activity-induced RV variation (30 m~s$^{-1}$ peak to peak, 19-day periodicity) prevents detection of additional planets &\citet{Kosiarek}\\
WASP-47&	Y	&	2 additional transiting planets with a 4-day giant planet; 1 non-transiting giant planet on 600-day orbit &\citet{Vanderburg47}\\
K2-106&	Y	&	1 additional non-transiting planet on 13-day orbit &\citet{Guenther}\\
55 Cnc&	Y	&	4 additional non-transiting planets between 14 and 5000 days including a close-in giant planet &\citet{Bourrier}\\
HD 213885&	Y	&	1 additional non-transiting planet on 5-day orbit &\citet{Espinoza}\\
\hline
K2-100&	N	&	Activity-induced RV variation (100 m~s$^{-1}$ peak to peak, 4.3-day periodicity) prevents detection of additional planets &\citet{Barragan100}\\
LTT 9779&	N	&	20 days of RV baseline &\citet{Jenkins}\\
TOI-849&	N	&	>400 days of RV baseline &\citet{Armstrong849}\\
\enddata
\end{deluxetable*}

\subsection{Implications for Planet Formation}
The top contenders of USP formation theory invoke the secular interaction between USPs and longer-period planets \citep{Petrovich,Pu}. Consider a USP that formed initially on longer orbital periods 2-10 days range. This is beyond the dust sublimation radius and where many {\it Kepler} planets are found today. If there are other planets in the same system with enough angular momentum deficit (AMD)  i.e. the angular momentum difference between a system with coplanar, circular orbits and a systems with eccentric, mutually inclined orbits, secular interaction between the planets could shuffle the AMD around. Since the progenitors of USPs are the innermost planet of their system, they have the lowest angular moment per unit mass. The same AMD could thus induce a significantly eccentric and inclined orbit. The augmented tidal interaction with the host star at periastron could then shrink the orbit of the USPs. This secular scenario has gained observational support as the USPs are indeed observed on more inclined orbits compared to other {\it Kepler} planets and USPs often have a longer orbital period ratio relative to their neighbors---likely a consequence of orbital decay  \citep{Dai2018,Steffen2016}. Here, we point out another observational support of the secular theory i.e. USP must have additional planets to initiate the secular interaction and provide enough AMD. In Table 3 we showed that USPs almost always have additional planetary companions (14/17). For 55 Cnc, the existing RV dataset is big enough to reveal orbital eccentricities of the various planets thereby allowing us to gauge whether the amount of AMD in a system could alter USP orbit significantly. Indeed, \citet{HansenZink} showed that the architecture of 55 Cnc does contain the requisite AMD and is in general consiten with the secular formation scenario. Future extensive RV follow-up of USPs may extend this AMD test to other systems.

Hot Neptunes, including those on sub-day orbits, show striking similarities with hot Jupiters: 1)they favor metal-rich environments; 2)they tend to be the only planet in a system; 3)their occurrence rate sums up to about 1\% occurrence rate within 10-days around sun-like stars \citep{Dong}. Another similarity between hot Jupiters and hot Neptune is that many hot Neptunes were also observed to be on misaligned orbits characterized by large stellar obliquities e.g. Kepler-63 \citep{Sanchis63}, HAT-P-11 \citep{Sanchis_hat} and WASP-107 \citep{Dai_107,Rubenzahl}. For hot Jupiters, a large spin-orbit angle has traditionally been interpreted as a signpost of a dynamically hot formation scenario that tilted planet orbit while also triggering orbital decay \citep[[see the view by][]{Dawson}. The large obliquity of many hot Neptunes is suggestive of a formation channel similar to that of the hot Jupiters. It will be instructive to extend obliquity measurements to ultra-hot Neptunes such as K2-100, LTT 9779 and TOI-849, although this is technically challenging with current generation spectrographs. \citet{Berger_age} also presented evidence that hot Neptunes are preferentially found around evolved stars. This could be an indication that the hot Neptunes migrated as a result of host stellar evolution.

Two recent planet formation theoretical works may also shed light on the distinction between USPs and hot Neptunes. \citet{Adams} performed a simple analysis that optimized the total energy of a pair of forming planets assuming the conservation of angular momentum, constant total mass reservoir and fixed orbital spacings. They found that when the total available mass is low ($\lesssim40M_\oplus$, a number that depends on stellar mass, semi-major axis etc), the energy-optimized state is an equal partition of mass between two competitively growing planets. This thus tends to create a multi-planet systems with intra-system uniformity that was seen in many observed sub-Neptune multi-planet systems \citep{Millholland,Weiss,Songhu}. On the other hand, when the total mass is high enough, the system switches to a different optimized state in which the mass is concentrated in one of the planets, thereby creating a dominant, possibly lonely planet that may undergo runaway accretion. Whether a system does go into the runaway accretion state, as pointed out by \citet{Lee_accretion} and \citet{Chachan}, further depends on the local hydrodynamic flow conditions, the local opacity and the time of core emergence i.e. whether gas is still present in the disk. These theories put the USPs and hot Neptunes into perspective: in a metal-rich (high [Fe/H]) disk,  the planetesimals assemble quickly and grow large enough to accrete the remaining gas in the disk \citep[see also][]{Dawson_met}. If the total planetesimal mass is above some threshold, one particular planet embryo grow to be the dominant planet in a system as \citet{Adams} would predict. In short, metal-rich systems tend to breed hot Neptunes and hot Jupiters. On the other hand, the rocky USP planets could be the product of late assembly in gas-depleted disk in which core assemebly proceeds oligarchically and slowly. This explains why their occurrence does not correlate strongly with host star metallicity. \citet{Adams} would further predict that USP planets come in multi-planet systems where the planets are similar in size. 

\begin{figure*}
\begin{center}
\includegraphics[width = .95\columnwidth]{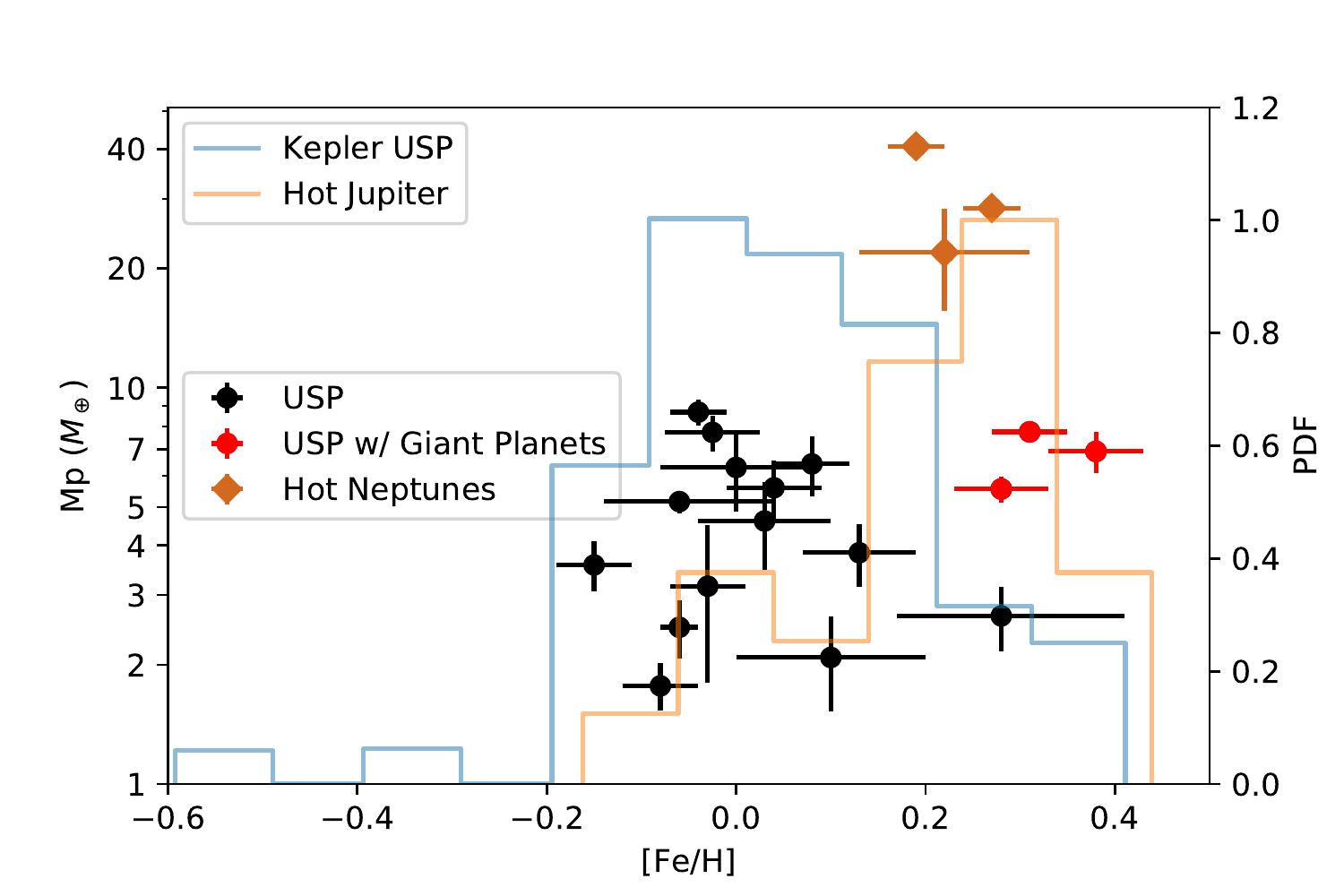}
\includegraphics[width = .95\columnwidth]{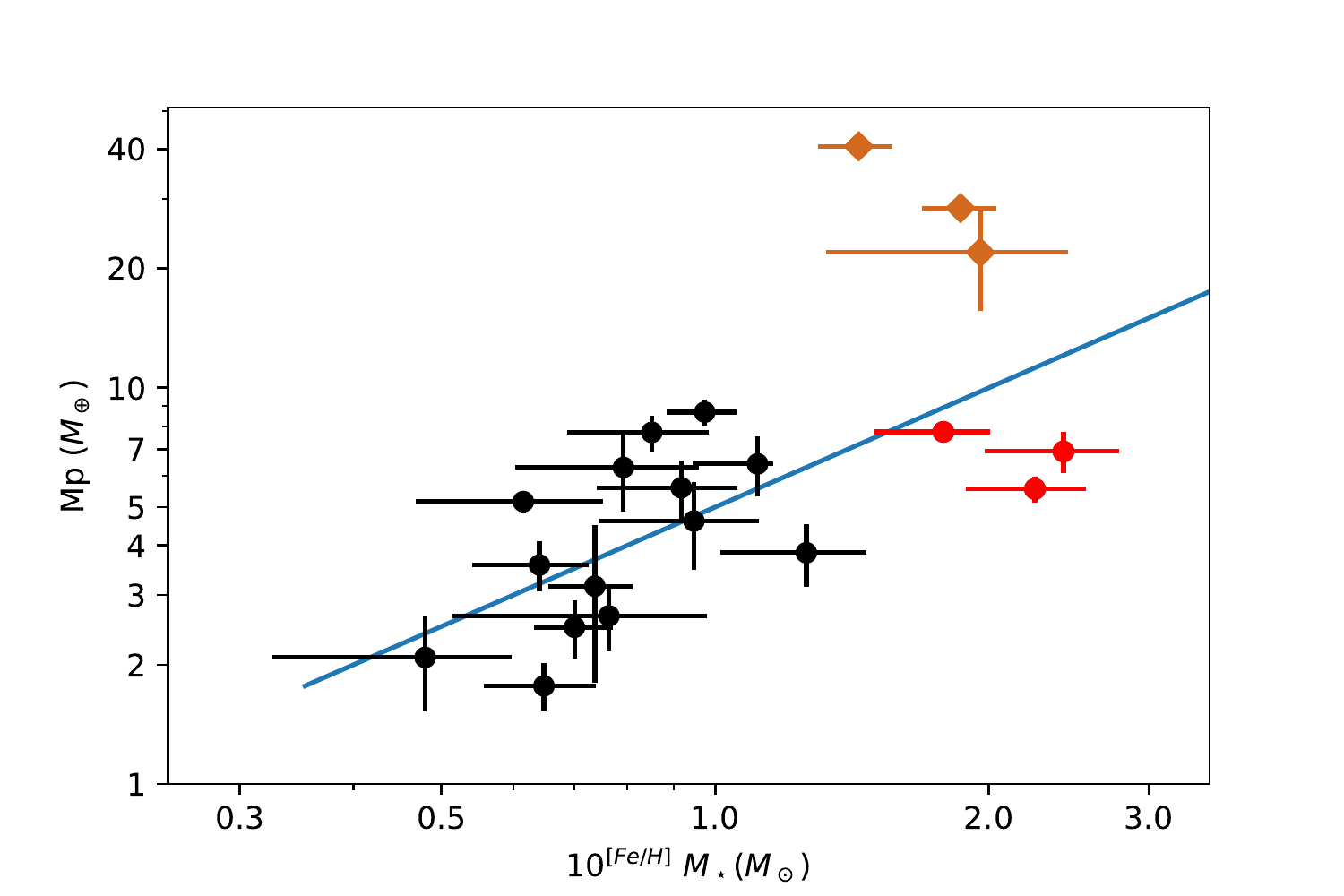}
\includegraphics[width = .95\columnwidth]{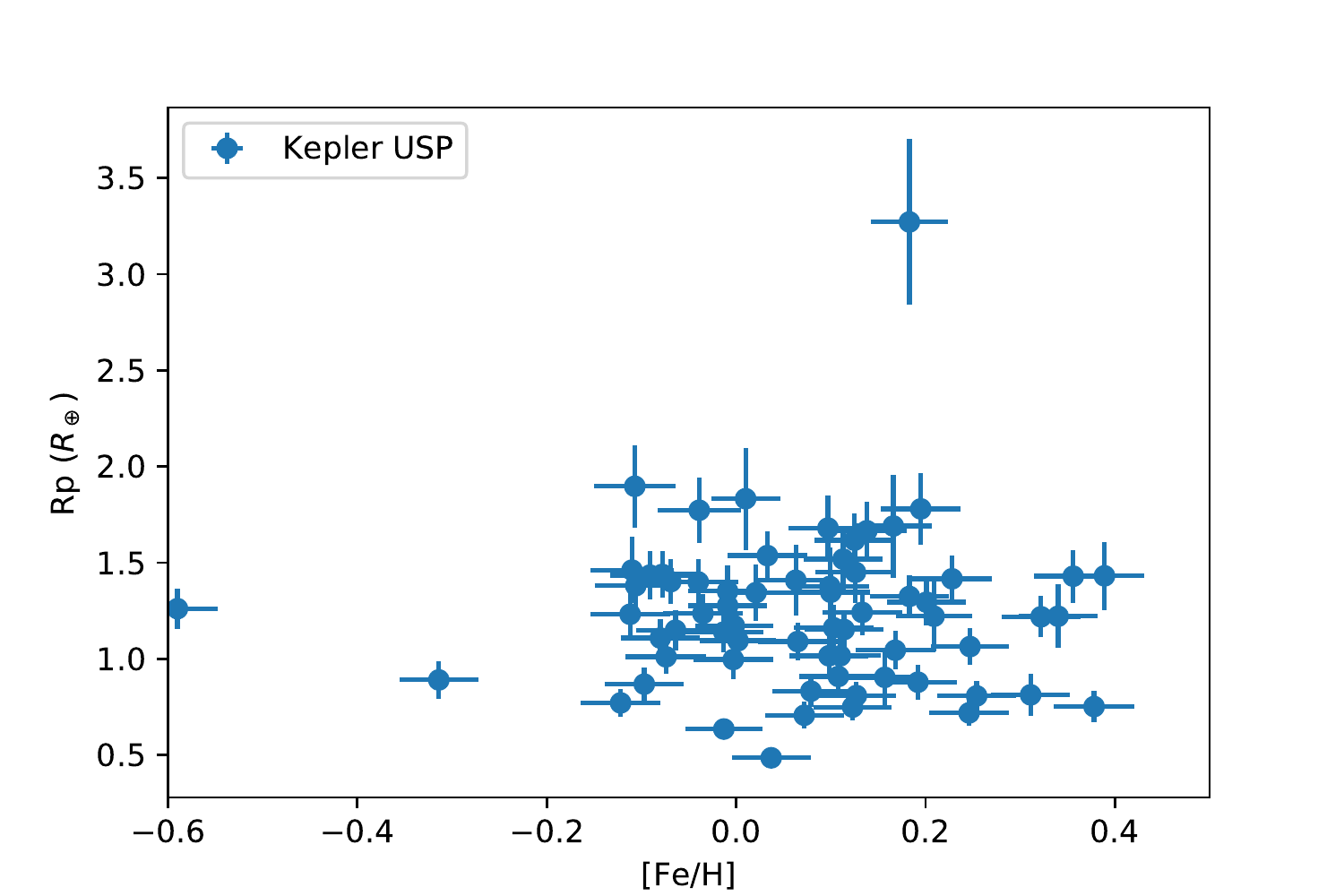}
\caption{Top left: the metallicity versus planet mass of USPs (black) and ultra-hot Neptunes (brown) with mass and radius measurements.The histograms the metallicity distribution of all {\it Kepler} USP (blue) and hot Jupiters (orange). Top-right: a proxy for disk solid density $10^{[Fe/H]}M_\star$ versus the planet mass of USPs (black) and ultra-hot Neptunes (brown). The implicit assumption is that protoplanetary disk density scales linearly host star mass and solid surface density scales with host star metallicity. A statistically strong (p = 0.04) correlation is observed. The blue line shows a linear scaling between $10^{[Fe/H]} M_\star$ and $M_p$. Bottom:  The  radii and host metallicity of the uniformly studied USP sample from {\it Kepler} \citep{USP}. Only 1 out of 67 is in the Neptune-radius regime (KOI-3913), suggesting that ultra-hot Neptunes are likely much rarer than USPs. Also note KOI-3913's relatively high metallicity similar to the confirmed ultra-hot Neptunes.}
\label{fig:metallicity}
\end{center}
\end{figure*}

Table 3 shows that there are three USPs that also have detected giant planet companions. They are 55 Cnc with a 14.6-day giant planet \citep{Butler55}; WASP-47 with with 4.2 and 600-day giant planets\citep{Becker47,Neveu} and HD 80653 with a strong RV trend likely indicative of a giant planet 0.37$\pm0.08$ $M_{\rm Jup}$(a$/$AU)$^{-2}$ \citep{Frustagli}. The attentive reader might have already guessed that these three systems are also the most metal-rich among the USP sample with [Fe/H] = 0.31$\pm0.04$, 0.38$\pm$0.05  and 0.28$\pm$0.05 respectively (red points in Fig. \ref{fig:metallicity}). These three USPs also happen to be the more massive ones among the USP sample (Fig. \ref{fig:metallicity}). One is tempted to ask: are USPs in metal-rich environments also the more massive ones? We use the $M_\star*10^{[Fe/H]}$ as a proxy for the surface density of solid material available to USP growth. \citet{Dai_mmen} applied a Minimum-mass Extrasolar Nebula framework to {\it Kepler} sub-Neptune planets which suggeested that the disk solid mass displayed a linear scaling with host star mass even within the innermost 1AU. $M_\star*10^{[Fe/H]}$ may be a reasonable proxy for the solid disk density. We found $M_\star*10^{[Fe/H]}$ and USP mass do show a positive correlation with a p-value of 0.04 in a Pearson test (Fig. \ref{fig:metallicity}). This correlation may suggest that the surface density of solids in the disk directly controls the overall availability of solids, the assembly rate, and eventually manifests as the size of planetary cores that emerge out of the disks.

\subsection{Threshold for runaway accretion?}     
As we argued above, USPs are probably a distinct group compared to hot Jupiters and hot Neptunes given the differences in host star metallicity and planet architecture. USPs are most likely cores of planets that escaped runaway accretion in the first place i.e. superEarths or sub-Neptunes. The rate of atmospheric erosion particularly photoevaporation shows a steep dependence on planet mass. It is suppressed by orders of magnitude for giant planets \citep[e.g.][]{Murray-Clay}. Briefly, this is because the gravitational potential well deepens with planet mass, but the heating efficiency of XUV irradiation does not. For example, the H/He envelopes on a $5\,M_\oplus$ planet can be easily stripped in 
100 Myr at 0.1 AU around a G-type star, but the mass loss timescale quickly shoots up to more than a Hubble time for planetary cores with  >10-15 $M_\oplus$ depending on the insolation the planet receives \citep{WangDai}. In summary, planets heavier than Neptune can only lose a small fraction of their envelope over their lifetime; thus they are unlikely to be stripped down to a rocky core that is observed as a USP planet today.

Therefore, the sample of USP planets help to place an upper limit on the threshold of runaway accretion. RV follow-up observations are heavily biased towards the detection of more massive planets. However, this bias works in our favor as we are probing an upper limit in mass. If one assumes a radiative outer layer, the critical mass for runaway accretion only has a
logarithmic dependence on the local disk properties  \citep[e.g.][]{Rafikov}. In other words, it is not sensitive to the location where the planets formed; and is estimated to be about $10\,M_\oplus$  \citep{Rafikov}. If one more carefully accounts for the envelope opacity, the threshold mass may be more variable between 2 and $8\,M_\oplus$ \citep{Lee2016}. The lower the local gas opacity, the lower the mass threshold for run-away accretion \citep[see also ][]{Chachan}. In this framework, the super-puffs-- low density, gas-rich planets with <5$M_\oplus$, >5 $R_\oplus$,  likely formed beyond the snowline where the opacity is low \citep[e.g. Kepler-51][]{Libby-Roberts}. Coming back to the current sample of 17 USPs, they are mostly consistent with exposed rocky cores. They seem to probe the upper limit of the threshold for run-away accretion. This seems to be at the predicted $8\,M_\oplus$ for the high-opacity inner disk ($\sim$ 0.1 AU) where USPs have likely formed. We note that the only USP that is suspected to have some atmosphere is 55 Cnc e \citep{Angelo} right at $\sim 8 M_\oplus$. This putative current-day atmosphere on 55 Cnc e was also suggested to be secondary in nature. 55 Cnc e was neither detected in Lyman-$\alpha$ \citep{Ehrenreich} or metastable He \citep{Zhang} transit observations.

\section{Summary}
In this work, we report the transiting USP TOI-1444b discovered in {\it TESS} photometry. We performed extensive follow-up of the system including AO imaging with Gemini/NIRI, Doppler monitoring with Keck/HIRES etc. These follow-up observations confirmed the planetary nature of the transiting signal and characterized the system. We also make use of this opportunity to analyze all USPs with precise mass and radius measurements and compare them with the recently reported ultra-hot Neptune planets. Our key findings are as follows:

\begin{itemize}
    \item {TOI-1444b is a 0.47-day transiting USP with a radius of 1.397$\pm0.064R_\oplus$ and a mass of 3.87$\pm0.71M_\oplus$ consistent with a rocky composition.}
    
    \item {We report a tentative 2.5$\sigma$ detection of the phase curve variation and secondary eclipse of TOI-1444b in {\it TESS} band. Future observation of this target in 2 sectors during the TESS Extended Mission (S49 and S52) could help confirm this detection. However, observations at different wavelengths are needed to disentangle the thermal emission and reflected light contribution.}
    
    \item {No additional transiting planets were found in the observed 9 sectors of {\it TESS} light curve. RV follow-up of Keck/HIRES revealed a non-transiting planet candidate on a 16-day orbit with a minimum mass of 11.8$\pm2.9M_\oplus$.}

    \item {USP planets and ultra-hot Neptunes are likely distinct groups of planets. Hot Neptunes preferentially occur around metal-rich systems, whereas USPs occur around stars with solar-like abundance. USPs are almost always found in multi-planet systems, whereas hot Neptunes tend to be ``lonely''. USPs are exposed rocky cores that cluster around Earth-like composition. Hot Neptunes require H/He atmosphere or other volatile to explain their mass and radius. ultra-hot Neptunes are likely rarer than USPs by 1-2 orders of magnitude.}
    
    \item{Metal-rich environments breed a diversity of planetary systems from USP, hot Neptunes to hot Jupiters \citep[see also][]{Petigura}. USPs in systems with more solid materials tend to be more massive, although none of them are above the $8M_\oplus$ threshold for runaway accretion.}
\end{itemize}

\software{{\sc AstroImage} \citep{Collins:2017}, {\sc Isoclassify} \citep{Huber}, {\sc MIST} \citep{MIST}, {\sc SpecMatch-Syn} \citep{Petigura_thesis}, {\sc Batman} \citep{Kreidberg2015}, {\sc emcee} \citep{emcee}, {\sc RVSearch}, {\sc RadVel} \citep{RadVel}}

\acknowledgements
We thank Heather Knutson, Yayaati Chachan and Shreyas Vissapragada for insightful discussions. We thank the time assignment committees of the University of California, the California Institute of Technology, NASA, and the University of Hawaii for supporting the TESS-Keck Survey with observing time at Keck Observatory and on the Automated Planet Finder.  We thank NASA for funding associated with our Key Strategic Mission Support project.  We gratefully acknowledge the efforts and dedication of the Keck Observatory staff for support of HIRES and remote observing.  We recognize and acknowledge the cultural role and reverence that the summit of Maunakea has within the indigenous Hawaiian community. We are deeply grateful to have the opportunity to conduct observations from this mountain.  We thank Ken and Gloria Levy, who supported the construction of the Levy Spectrometer on the Automated Planet Finder. We thank the University of California and Google for supporting Lick Observatory and the UCO staff for their dedicated work scheduling and operating the telescopes of Lick Observatory.  This paper is based on data collected by the TESS mission. Funding for the TESS mission is provided by the NASA Explorer Program. This work makes use of observations from the LCOGT network. LCOGT telescope time was granted by NOIRLab through the Mid-Scale Innovations Program (MSIP). MSIP is funded by NSF. Part of this work has been carried out within the framework of the National Centre of Competence in Research PlanetS supported by the Swiss National Science Foundation. ECM acknowledges the financial support of the SNSF. Based on observations obtained at the international Gemini Observatory, a program of NSF’s NOIRLab, which is managed by the Association of Universities for Research in Astronomy (AURA) under a cooperative agreement with the National Science Foundation. on behalf of the Gemini Observatory partnership: the National Science Foundation (United States), National Research Council (Canada), Agencia Nacional de Investigaci\'{o}n y Desarrollo (Chile), Ministerio de Ciencia, Tecnolog\'{i}a e Innovaci\'{o}n (Argentina), Minist\'{e}rio da Ci\^{e}ncia, Tecnologia, Inova\c{c}\~{o}es e Comunica\c{c}\~{o}es (Brazil), and Korea Astronomy and Space Science Institute (Republic of Korea). J.M.A.M. is supported by the National Science Foundation Graduate Research Fellowship Program under Grant No. DGE-1842400. J.M.A.M. acknowledges the LSSTC Data Science Fellowship Program, which is funded by LSSTC, NSF Cybertraining Grant No. 1829740, the Brinson Foundation, and the Moore Foundation; his participation in the program has benefited this work. D.H. acknowledges support from the Alfred P. Sloan Foundation, the National Aeronautics and Space Administration (80NSSC19K0379), and the National Science Foundation (AST-1717000). Resources supporting this work were provided by the NASA High-End Computing (HEC) Program through the NASA Advanced Supercomputing (NAS) Division at Ames Research Center for the production of the SPOC data products. We acknowledge the use of public TESS Alert data from pipelines at the TESS Science Office and at the TESS Science Processing Operations Center. JNW thanks the Heising-Simons foundation for support. D. D. acknowledges support from the TESS Guest Investigator Program grant 80NSSC19K1727 and NASA Exoplanet Research Program grant 18-2XRP18\_2-0136.

\bibliography{main}

\begin{deluxetable}{ccccc}
\tabletypesize{\scriptsize}
\label{tab:rv}
\tablecaption{Keck/HIRES Radial Velocities }
\tablehead{
\colhead{Time (BJD)} & \colhead{RV (m/s)} & \colhead{RV Unc. (m/s)} & \colhead{$S_HK$}& \colhead{$S_HK$ Unc.} }
\startdata
2458832.753061&0.89&2.34&0.143&0.001\\
2459013.084714&6.11&1.72&0.147&0.001\\
2459071.083021&-0.30&1.46&0.146&0.001\\
2459072.076353&-4.44&1.28&0.149&0.001\\
2459073.046224&-5.81&1.32&0.147&0.001\\
2459077.802623&-2.49&1.21&0.148&0.001\\
2459077.909326&1.18&1.20&0.149&0.001\\
2459078.064631&0.49&1.32&0.149&0.001\\
2459086.775934&-2.07&1.78&0.092&0.001\\
2459087.847993&2.15&1.67&0.148&0.001\\
2459088.799006&3.13&1.44&0.147&0.001\\
2459088.919843&2.77&1.51&0.133&0.001\\
2459089.048409&3.10&1.33&0.143&0.001\\
2459089.747648&10.33&1.19&0.147&0.001\\
2459089.983381&-0.52&1.32&0.146&0.001\\
2459090.039667&1.44&1.23&0.147&0.001\\
2459090.746892&9.28&1.14&0.145&0.001\\
2459090.987475&-4.59&1.31&0.142&0.001\\
2459091.06383&0.75&1.31&0.144&0.001\\
2459091.745834&5.47&1.21&0.144&0.001\\
2459091.925965&4.26&1.27&0.133&0.001\\
2459092.070412&3.45&1.19&0.135&0.001\\
2459092.740714&5.19&1.22&0.146&0.001\\
2459092.878645&-1.57&1.37&0.145&0.001\\
2459093.045547&1.95&1.35&0.143&0.001\\
2459094.745288&-6.12&1.19&0.143&0.001\\
2459094.973684&-2.37&1.26&0.144&0.001\\
2459097.887255&-1.04&1.28&0.148&0.001\\
2459098.03801&-11.51&1.49&0.140&0.001\\
2459099.883517&-7.92&1.24&0.138&0.001\\
2459101.020237&-1.92&1.38&0.143&0.001\\
2459101.785447&-3.98&1.20&0.144&0.001\\
2459102.016817&3.26&1.27&0.147&0.001\\
2459114.738603&5.24&1.23&0.147&0.001\\
2459114.969879&-9.34&1.31&0.145&0.001\\
2459115.761093&3.02&1.32&0.147&0.001\\
2459115.977161&-3.31&1.31&0.146&0.001\\
2459116.956212&-1.33&1.37&0.145&0.001\\
2459117.738884&-7.76&1.26&0.149&0.001\\
2459117.782287&-3.76&1.26&0.146&0.001\\
2459117.995658&0.04&1.43&0.143&0.001\\
2459118.718154&-4.30&1.30&0.148&0.001\\
2459118.951548&7.70&1.35&0.140&0.001\\
2459119.745284&-1.74&1.31&0.146&0.001\\
2459119.876639&4.72&1.41&0.144&0.001\\
2459120.837472&2.60&1.29&0.145&0.001\\
2459120.970506&0.38&1.52&0.134&0.001\\
2459121.742505&2.08&1.44&0.147&0.001\\
2459121.952438&1.64&1.30&0.142&0.001\\
2459122.725583&5.83&1.33&0.147&0.001\\
2459122.945804&-0.12&1.58&0.146&0.001\\
2459123.717734&9.37&1.31&0.146&0.001\\
2459123.897828&-1.54&1.42&0.144&0.001\\
2459142.953366&-7.59&1.87&0.145&0.001\\
2459151.793062&-9.99&1.89&0.137&0.001\\
2459189.785436&0.05&2.05&0.149&0.001\\
\enddata
\end{deluxetable}

\end{document}